\documentclass[structabstract]{aa}  
\usepackage{graphicx}
\usepackage{txfonts}
\begin{document}
   \title{Very low mass white dwarfs with a C-O core}


   \author{P. G. Prada Moroni
          \inst{1,2}
          \and
          O. Straniero
          \inst{3}
                    }

   \institute{Physics Department ``E. Fermi'', University of Pisa,
largo B. Pontecorvo 3, I-56127, Pisa, Italy\\              
         \and
             INFN, largo B. Pontecorvo 3, I-56127, Pisa, Italy\\
             \email{prada@df.unipi.it}
         \and
             INAF -- Osservatorio Astronomico di Collurania, via Maggini, I-64100, Teramo, Italy\\
             \email{straniero@oa-teramo.inaf.it}
             }

   \date{Received July 8, 2009; accepted September 9, 2009}

 
  \abstract
   {The lower limit for the mass of white dwarfs (WDs) with
C-O core is commonly assumed to be roughly 0.5 M$_{\odot}$. As a
consequence, WDs of lower masses are usually identified as He-core remnants.} 
   {However, when the initial mass of the progenitor star is in between 1.8 and 3 M$_\odot$,
   which corresponds to the so called red giant (RGB) phase transition, 
 the mass of the H-exhausted core at the tip of the RGB is 0.3$<$M$_H$/M$_{\odot}$$<$0.5.
  Prompted by this well
     known result of stellar evolution theory, we
     investigate the possibility to form C-O WDs with mass M $<$ 0.5 M$_{\odot}$.}
   {The pre-WD evolution of stars with initial mass of about 2.3 M$_\odot$,  undergoing 
   anomalous mass-loss episodes during the RGB phase and leading to the formation of WDs with
   He-rich or CO-rich cores have been computed. The cooling sequences of the resulting 
   WDs are also described.}
   {We show that the minimum mass for a C-O WD is about 0.33 M$_{\odot}$,
     so that both He and C-O core 
    WDs can exist in the mass range 0.33-0.5 M$_{\odot}$. The models computed
    for the present paper provide the theoretical tools to indentify the
    observational counterpart of very low mass remnants with a C-O core among those
    commonly ascribed to the He-core WD population in the progressively
    growing sample of observed WDs of low mass. 
    Moreover, we show that the central He-burning phase of the stripped
    progeny of the 2.3 M$_\odot$ star lasts longer and longer as the total
    mass decreases. In particular, the M= 0.33 M$_{\odot}$ model takes about 800 Myr to 
    exhausts its central helium, which is more than three time longer  
    than the value of the standard 2.3 M$_{\odot}$ star: it is, by far, the
    longest core-He burning lifetime.
    Finally, we find the  occurrence of gravonuclear instabilities during
    the He-burning shell phase.}
   {}


   \maketitle
%

\section{Introduction}
The value for the minimum mass of white dwarfs (WDs) with a carbon-oxygen (C-O) core
is commonly agreed to be around 0.5 M$_{\odot}$ (Weideman \cite{weidemann},
Meng et al. \cite{meng}). 
The main reason for such a belief is a firm result of the theory of
stellar evolution and can be easily understood looking at figure \ref{MheTip},
 which shows the mass of the hydrogen exhausted core M$_H$
 at the first thermal pulse on the asymptotic giant (AGB, solid line) and at the tip of the
 red giant branch (RGB tip, dashed line) as a function of their initial mass. 
 As it is well known, for initial
 masses lower than 3 M$_{\odot}$, the value of M$_H$ at the first thermal pulse is nearly constant
 around 0.55 M$_{\odot}$, the exact value depending on the chemical
 composition. This value represents the minimum mass for C-O WDs produced by the evolution   
of single, low- or intermediate-mass, stars undergoing normal\footnote{e.g. a Reimers like mass loss rate 
during the RGB phase.} mass loss during the pre-AGB evolution. 
Nonetheless, anomalous mass loss episodes occurring on the RGB or during the core He-burning phase,
as caused by either  
Roche-lobe overflow in close-binary evolution or tidal stripping in close
encounters between stars in high density environments, could interrupt or change the 
stellar evolution, leading to the production of a smaller remnant.
Actually, slightly smaller C-O WDs results from the evolution of a low-mass star 
(M$<$1.5-1.7 M$_\odot$, the excact value depends on the composition), 
which loses its envelope before the onsent of the AGB, 
the so-called AGB-manqu\'e (Sweigart, Mengel \& Demarque
\cite{sweigart74}, Caloi \cite{caloi}, Greggio \& Renzini  \cite{greggio90},
Castellani  \& Tornamb\'e \cite{ct91}, Dorman, Rood \& O'Connell
\cite{dorman93}). 
Examples of these objects are the extreme horizontal branch stars
(EHB) in globular clusters (GCs) and the field subdwarf B stars, which are
core He-burning stars with extremely thin hydrogen envelopes that eventually
become C-O core WDs of low mass (Castellani
\& Castellani \cite{castellani93}, Castellani, Luridiana \& Romaniello
\cite{castellani94}, D'Cruz et al \cite{dcruz96}, Brown et
al. \cite{brown2001}, Han et al. \cite{han02}, Han et al. \cite{han03}, Castellani, Castellani \& Prada
Moroni \cite{castellani06}, Han \cite{han08}). 
But even in such a case, the theory of stellar evolution predicts the existence of a lower limit, since
the mass of the H-exhausted core at the He ignition is slightly lower than
 0.5 M$_{\odot}$ \footnote{The exact value depending on the chemical composition: 
from 0.46 to 0.50 M$_{\odot}$ for Z=0.04 to
Z=0.0001.} for these low-mass stars (Castellani, Chieffi \& Straniero
\cite{ccs92}, Dominguez et al. \cite{dominguez}, Girardi \cite{girardi},
Castellani et al. \cite{cast2000}). 
In practice, if the mass of the electron-degenerate He-rich core were lower,
the cooling processes, i.e. the electronic
thermal conduction and the plasma neutrino emission, would prevail on 
the heating caused by the release of gravitational energy, the
He burning would be skipped and a He-core WD would be produced.  

The scenario described above explains why WDs whose mass is lower than 
0.5 M$_\odot$ are commonly believed to have a He-rich core, but this is not the whole story. 
In fact, stars with initial mass in the range 1.8 to 3 M$_\odot$, which undergo intense 
and rapid mass-loss episode, could produce both CO- or He- rich WDs
with mass lower than 0.5 M$_\odot$. 
In these stars, the degree of electron degeneracy and, in turn, the cooling processes, 
in the core developed during the RGB are weaker and the He ignition occurs when its mass is lower 
(see e.g. figure 1).    
As early as 1985 Iben \& Tutukov, in a very  
instructive paper, showed that a star with an initial mass of 3 M$_{\odot}$
evolving in a close-binary system can produce a remnant of 0.4 M$_{\odot}$ with a sizeable C-O core.
More recently, Han, Tout \& Eggleton \cite{han}, obtain a similar result
by computing the evolution of close binary systems by selecting  different 
initial parameters. They showed, in particular, that a star with an initial mass M$_1$= 2.51
M$_{\odot}$, belonging to a binary with mass ratio q=M$_1$/M$_2$=2 and initial period
P=2.559 d, eventually succeeds to ignite the helium-burning, even if they did
not followed the evolution up to the formation of the C-O WD, 
since their code met numerical problems when the primary star was as low as 0.33
 M$_{\odot}$ with a C-O core of 0.11 M$_{\odot}$.

In this framework and by means of detailed computations performed with
a full Henyey code able to follow consistently the evolution of stars from the
pre-main sequence to the final cooling phase of WDs, we have investigated possible
 evolutionary scenarios leading to the production of very low mass C-O WDs
 (VLMWDs), that is with mass lower than $< 0.5$ M$_{\odot}$. 
The peculiar features of the corresponding cooling evolution will be also discussed.

\section{The red giant phase transition}
The stellar models showed in the present work
 have been computed with an updated version of
 FRANEC (Prada Moroni \& Straniero \cite{pgpm02}, Degl'Innocenti et al. \cite{deglinnocenti}),
 a full Henyey evolutionary code.  
We adopted a metallicity Z=0.04 and an initial helium abundance Y=0.32 
suitable for the very metal-rich stars belonging to some open galactic
cluster, such as NGC6791.
%
   \begin{figure}
   \vspace{0.7cm}
   \centering
    \includegraphics[width=0.85\linewidth]{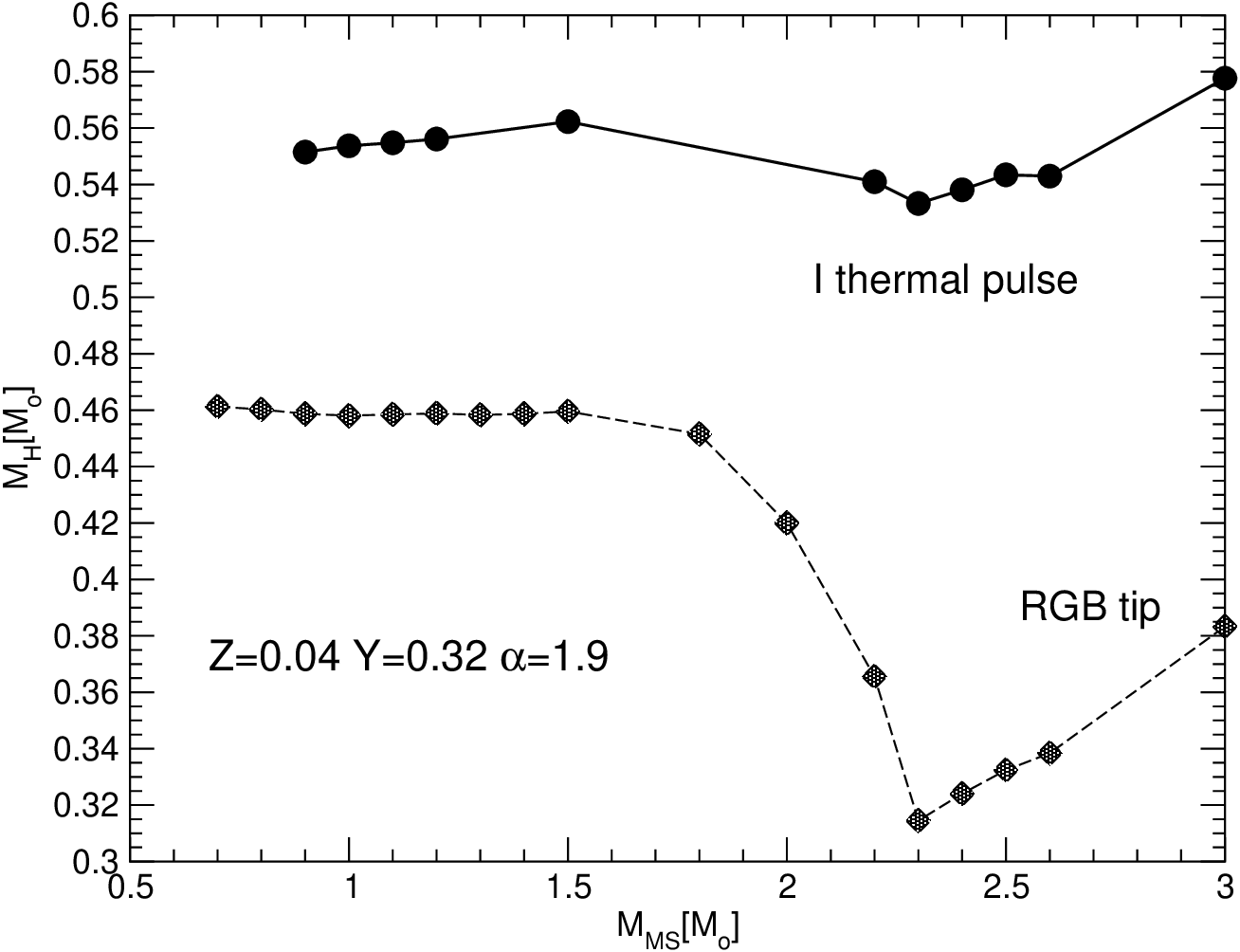}
    \caption{Mass of the hydrogen exhausted core M$_H$
              at the first thermal pulse on the AGB (circles and solid line) 
              and at the tip of the RGB (diamond and
             dashed line) as a function of the initial
             mass for stars with Z=0.04 Y=0.32. 
             }
    \label{MheTip}%
    \end{figure}
%
%
   \begin{figure}
   \vspace{0.7cm}
   \centering
    \includegraphics[width=0.85\linewidth]{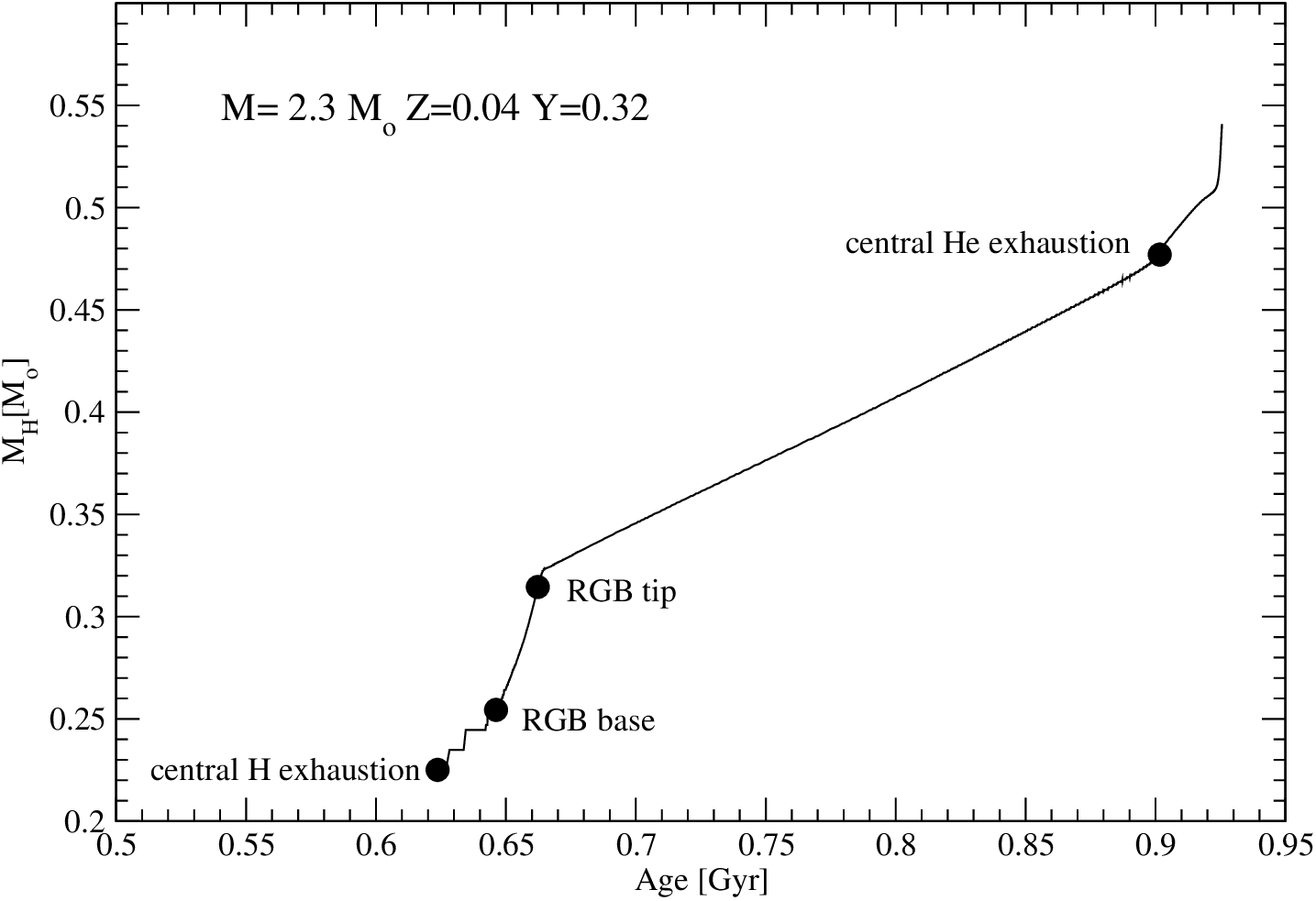}
    \caption{Mass of the hydrogen exhausted core as a function
             of the age for a star of M=2.3 M$_{\odot}$, Z=0.04 Y=0.32.}
    \label{MheAge2p3}%
    \end{figure}
%
A simple way to understand the findings by Iben \& Tutukov \cite{iben85}
 and Han, Tout \& Eggleton \cite{han} is to analyze figure \ref{MheTip}, 
which shows the dependence of M$_H$ at the first thermal pulse (solid line) and
at the RGB tip (dashed line) on the initial mass for isolated-single stars.
 At the beginning of the thermally pulsing AGB phase, M$_H$ is almost constant, around 0.55 M$_{\odot}$, 
for initial masses lower than 3 M$_{\odot}$, but at the tip of RGB
it presents a deep minimum around 2.3 M$_{\odot}$, for this particular chemical
 composition (Sweigart, Renzini and Greggio 1990; 
 Castellani, Chieffi, Straniero \cite{ccs92}, Bono et
 al. \cite{bono97b}). 
Such a behavior is the consequence of the physical conditions 
occurring in the helium-core at the ignition of the 3$\alpha$ nuclear reaction. 

For initial mass lower than 1.5-1.7 M$_{\odot}$, a regime of strong
electron-degeneracy develops in the helium core at the beginning of the red
giant phase, and the ignition of the 3$\alpha$ 
occurs in a violent nuclear runaway, the so-called helium-flash (Hoyle \&
Schwarzschild \cite{hoyle}) when M$_{H}$ approaches about
 0.46 M$_{\odot}$, for metal-rich stars, and 0.50 M$_{\odot}$, for metal-poor
stars. The first ignition point is off center and it is followed by a series of weaker 
and more central flashes, untill a quiescent He burning takes place in the centre.
From this point on, the larger the initial mass, the weaker the electron-degeneracy of
 the core in the red giant phase.  Consequently,  
 the mass of the H-exhausted core required for the 3$\alpha$ ignition is smaller and smaller. 
When the initial mass is as high as about 2.3 M$_{\odot}$, 
the helium burning ignites almost quiescently, through a single and central weak flash.
For higher masses, the core in the red giant phase is not degenerate
 any more. Nevertehless, for stars with initial mass larger than about 2.3 M$_\odot$,
 the larger the initial mass,
 the larger the M$_{H}$ at the tip of the RGB. In this stars, indeed, the helium in
 the core is mainly produced during the central hydrogen-burning phase, which
 occurs in a convective core whose mass extension is a growing function of the 
initial total mass. 
Such a variation of the physical conditions at the He ignition, 
gives rise to a minimum M$_{H}$ at the RGB tip that marks the transition 
between low and intermediate mass stars (the so called RGB phase transition,
 after Renzini \& Buzzoni \cite{renzini}).
 For the present composition (Z=0.04 Y=0.32), such a minimum occurs for
 M=2.3 M$_{\odot}$ and M$_H$=0.315 M$_{\odot}$. 
The RGB phase transition occurs at larger masses as the initial helium
    abundance decreases or the initial metallicity increases. Moreover, the
    value of the minimum M$_{H}$ at the RGB tip depends slightly on
    chemical composition (see e.g. Sweigart, Greggio \& Renzini \cite{sweigart89}).
Notice that, as previously stated, this minimum disappears at the 
first thermal pulse (see the upper curve in figure \ref{MheTip}). The reason is
that the H-burning shell continues to process hydrogen during the core He burning 
and move furtherly outward. Moreover,
 it is a well established result of the theory of stellar evolution that the
 lower the mass of the H-exhausted core at the onset of the 3$\alpha$, the fainter the 
star and the longer the central helium-burning phase. Thus, the H-burning shell
 will work for a longer
 time for stars belonging to the RGB phase transition. This is the reason why stars with
 initial mass lower than 3 M$_{\odot}$ enter the AGB phase with nearly the
 same M$_H$ (Dominguez et al. \cite{dominguez}). 
Since Iben \cite{iben67}, who provided the first detailed study of the
 evolutionary characteristics of stars belonging to the RGB phase transition, 
 many studies were focused on this particular range of stellar mass
 (see e.g. Sweigart, Greggio \& Renzini \cite{sweigart89}; Sweigart, Greggio \& Renzini
\cite{sweigart90}; Castellani, Chieffi \& Straniero \cite{ccs92}),
  but only a few dicussed the possible compact remnants of these evolutionary sequences.  

Figure \ref{MheAge2p3} illustrates the temporal evolution of the mass of the
hydrogen-exhausted core M$_H$ for a star with initial mass 2.3 M$_\odot$. 
The four circles mark the following evolutionary phases: the central-hydrogen exhaustion,
 the base and the tip of the RGB, respectively, and the central-helium exhaustion.
As one can easily see, when the star exhausts its central hydrogen, the mass 
of the core is already 0.225 M$_{\odot}$, which is more than the 70\% of
the value at the RGB tip. In fact, a sizeable convective core was developed
during the previous main sequence phase, which was supported by the H burning through the CNO cycle.
 When the star reaches the base of the RGB, the mass of the core is
 0.254 M$_{\odot}$, which is about the 80\% of the mass at the
 He ignition, namely 0.315 M$_{\odot}$.

This means that, since a large fraction of the helium-core required
 for the ignition of the He-burning is already developed at the beginning
 of the red giant, a star with mass of about 2.3
 M$_{\odot}$, that undergoes a strong mass loss episode during or immediately after
 the red giant phase can easily experience the core-He burning phase
 and produce a C-O VLMWD, that is a C-O WD with a mass substantially lower
 than 0.5 M$_{\odot}$, provided that the residual total mass after the
 anomalous mass loss episode is larger than about 0.32 M$_{\odot}$.
\section{Evolution of the stripped progeny of the 2.3 M$_{\odot}$ star}
%
    \begin{figure*}
     \centering
     \begin{minipage}[c]{0.5\linewidth}
       \vspace{0.9cm}
       \centering \includegraphics[width=0.8\linewidth]{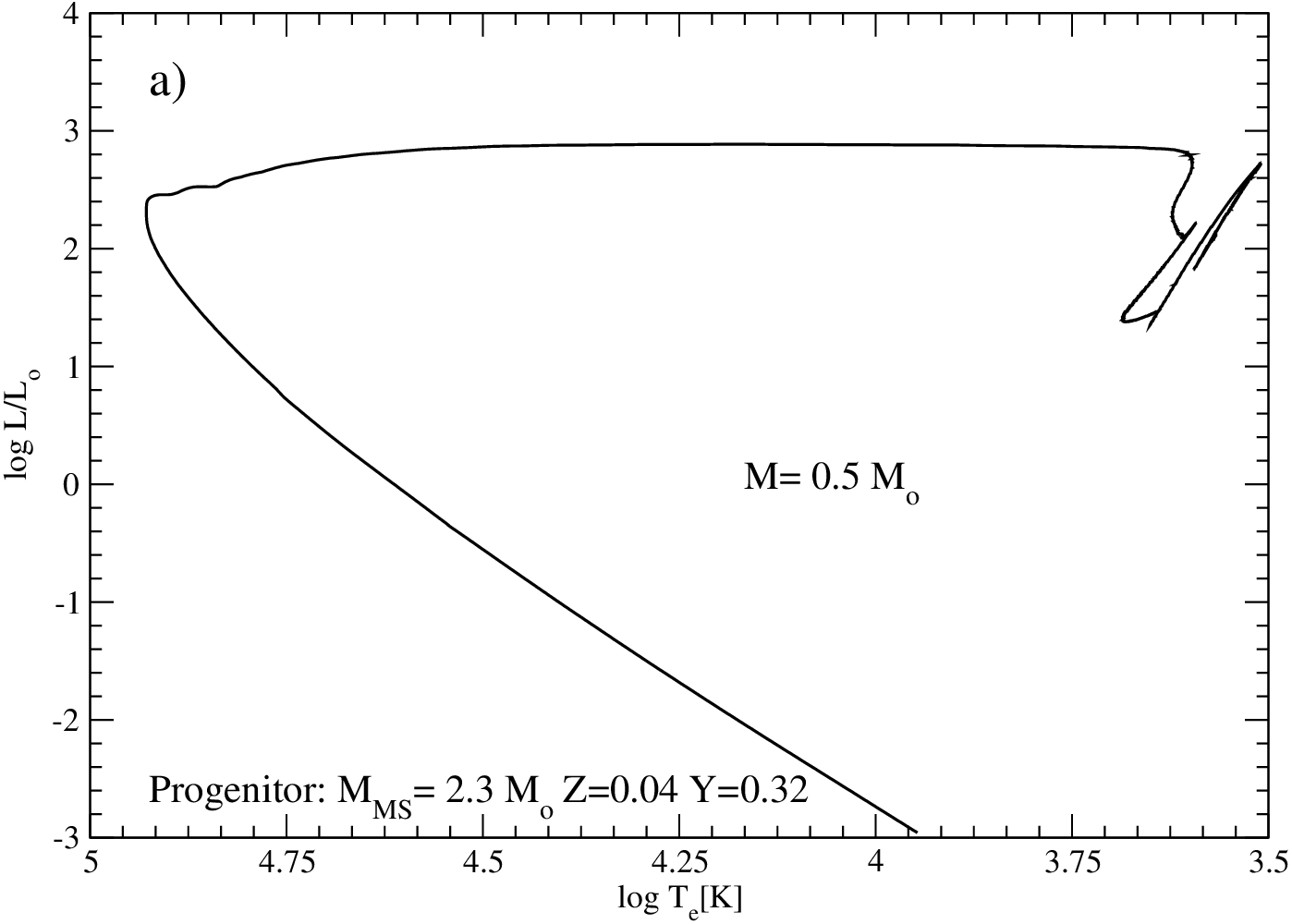}
     \end{minipage}%
     \begin{minipage}[c]{0.5\linewidth}
       \vspace{0.9cm}
       \centering \includegraphics[width=0.8\linewidth]{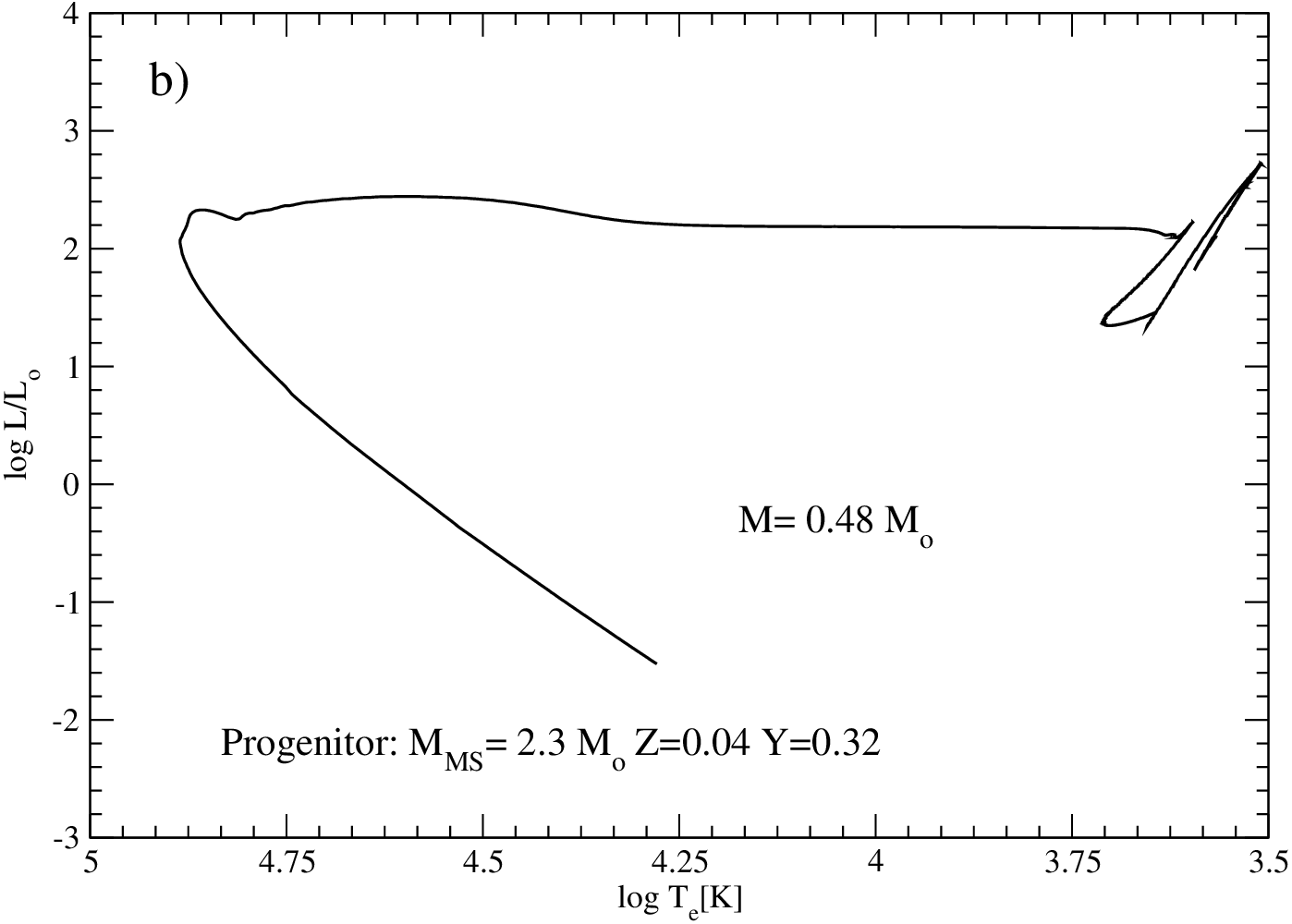}
     \end{minipage}
     \begin{minipage}[c]{0.5\linewidth}
       \vspace{0.9cm}
       \centering \includegraphics[width=0.8\linewidth]{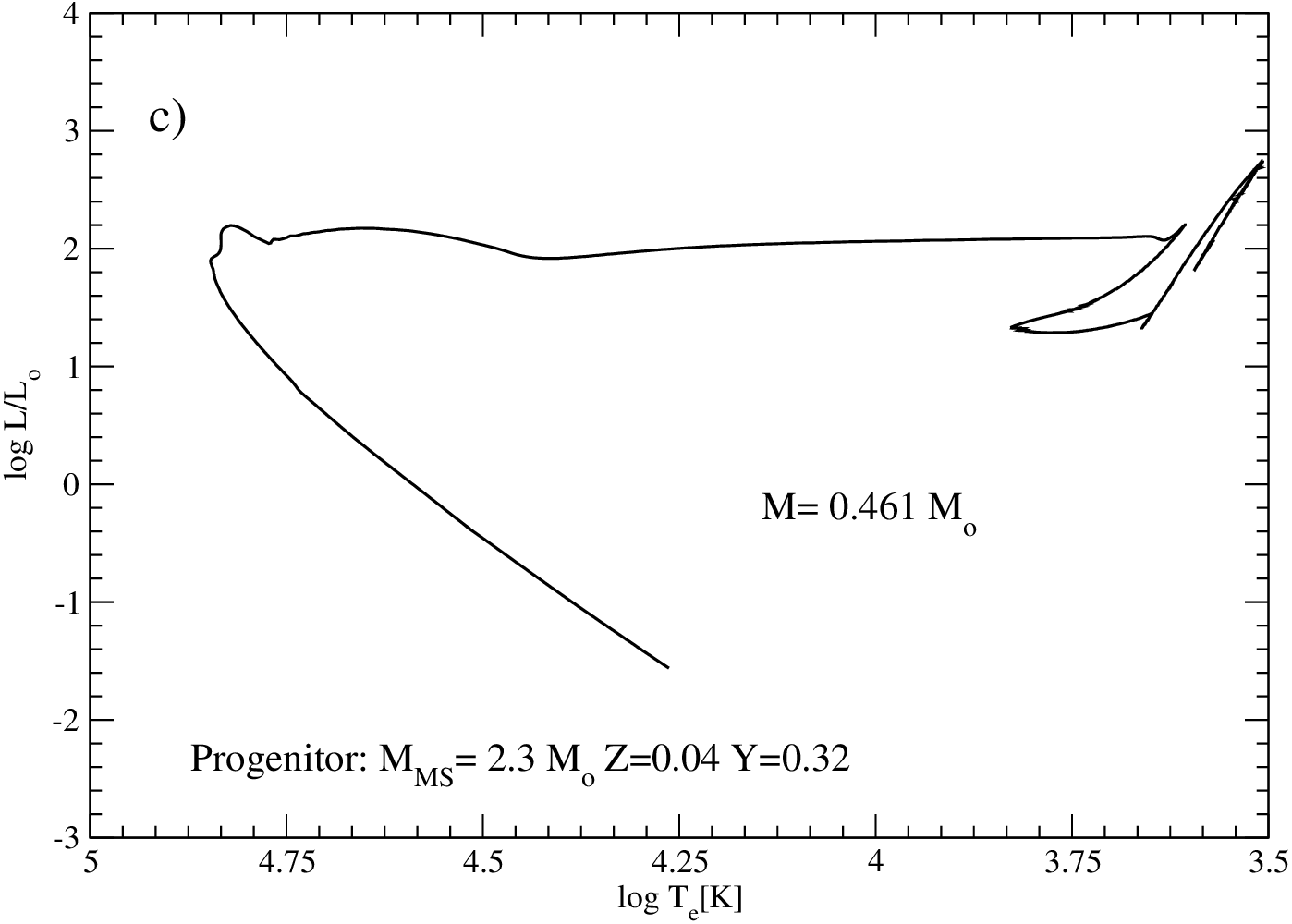}
     \end{minipage}%
     \begin{minipage}[c]{0.5\linewidth}
       \vspace{0.9cm}
       \centering \includegraphics[width=0.8\linewidth]{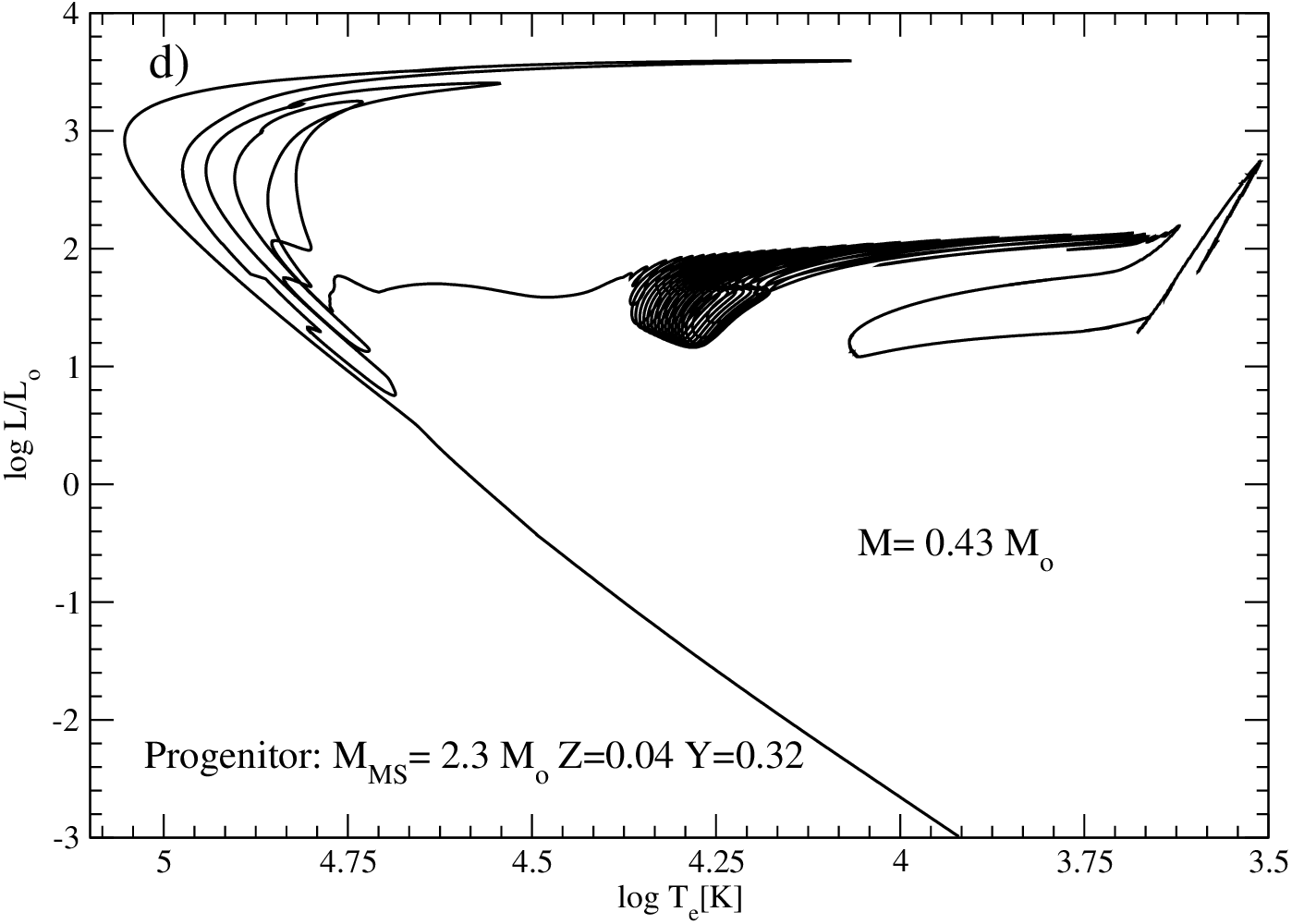}
     \end{minipage}
     \begin{minipage}[c]{0.5\linewidth}
       \vspace{0.9cm}
       \centering \includegraphics[width=0.8\linewidth]{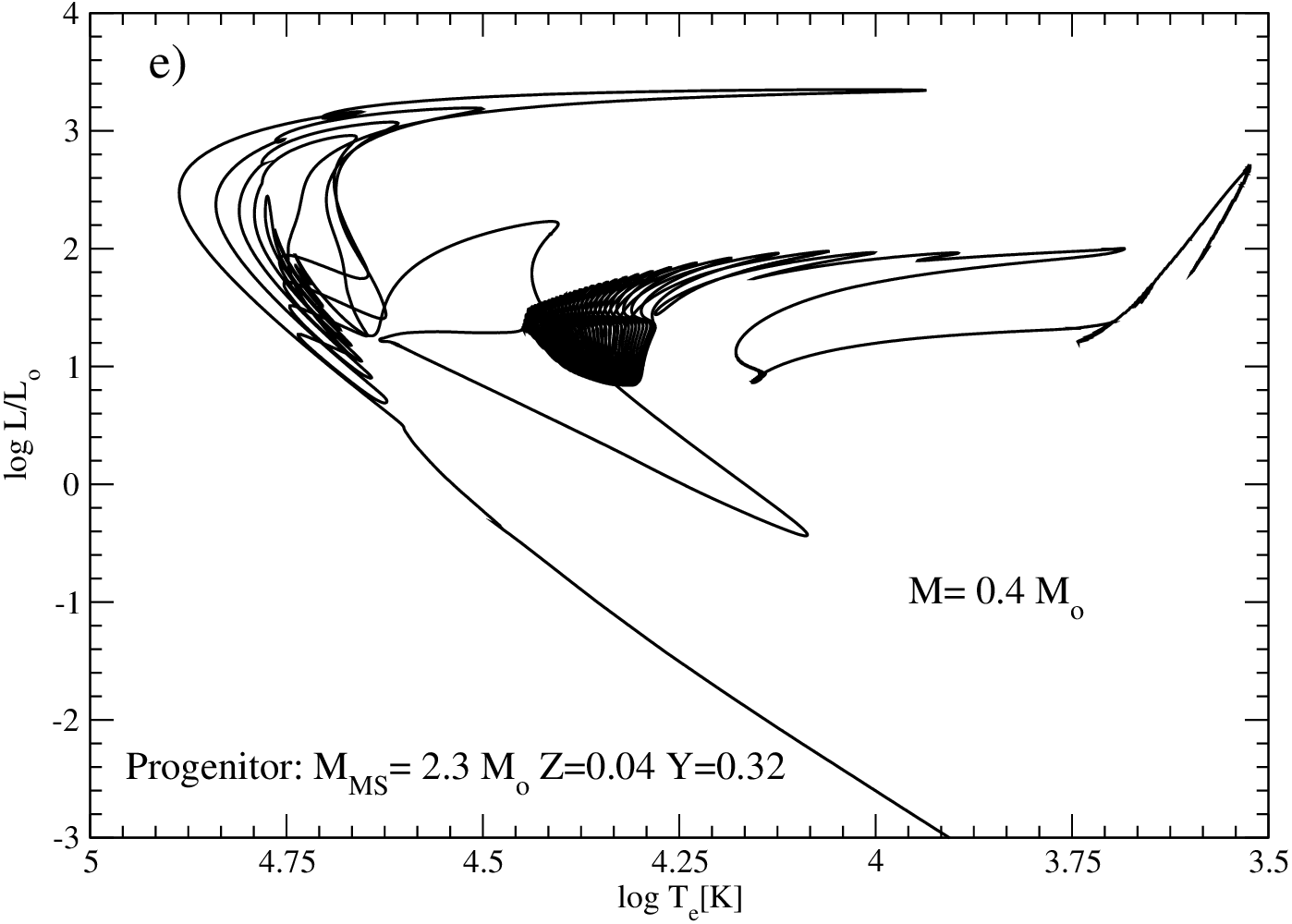}
     \end{minipage}%
     \begin{minipage}[c]{0.5\linewidth}
       \vspace{0.9cm}
       \centering \includegraphics[width=0.8\linewidth]{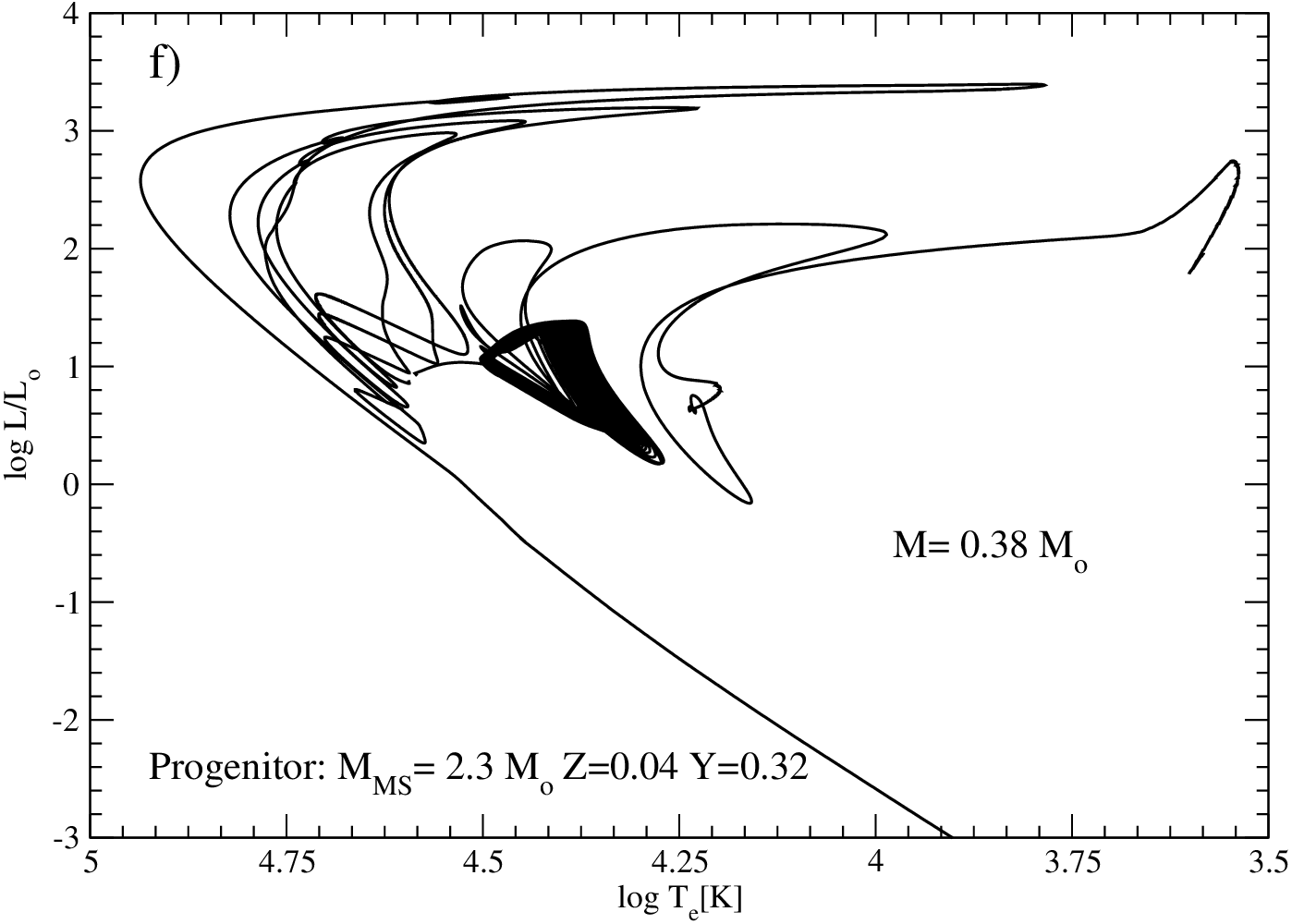}
     \end{minipage}
     \begin{minipage}[c]{0.5\linewidth}
       \vspace{0.9cm}
       \centering \includegraphics[width=0.8\linewidth]{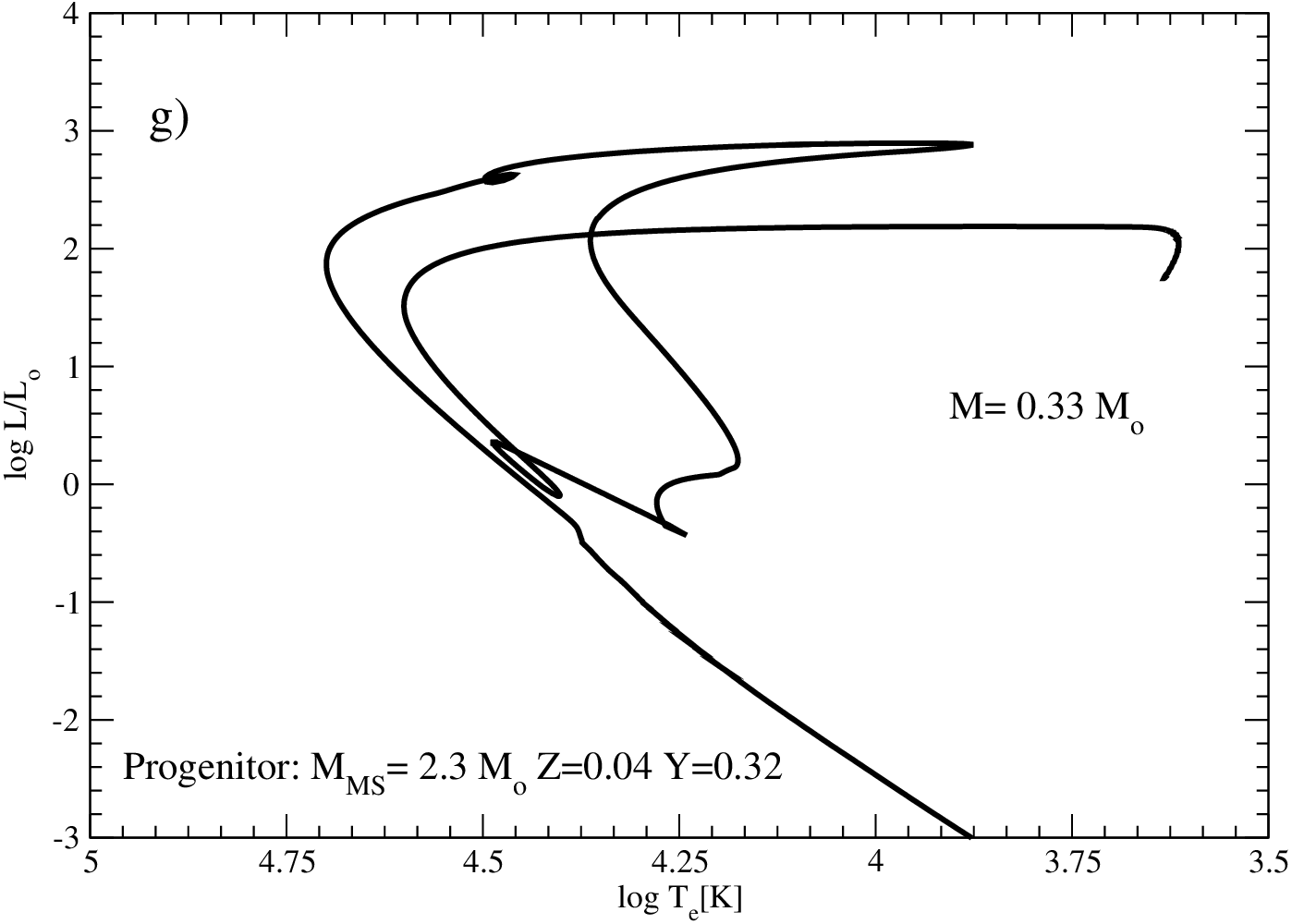}
     \end{minipage}%
     \begin{minipage}[c]{0.5\linewidth}
       \vspace{0.9cm}
       \centering \includegraphics[width=0.8\linewidth]{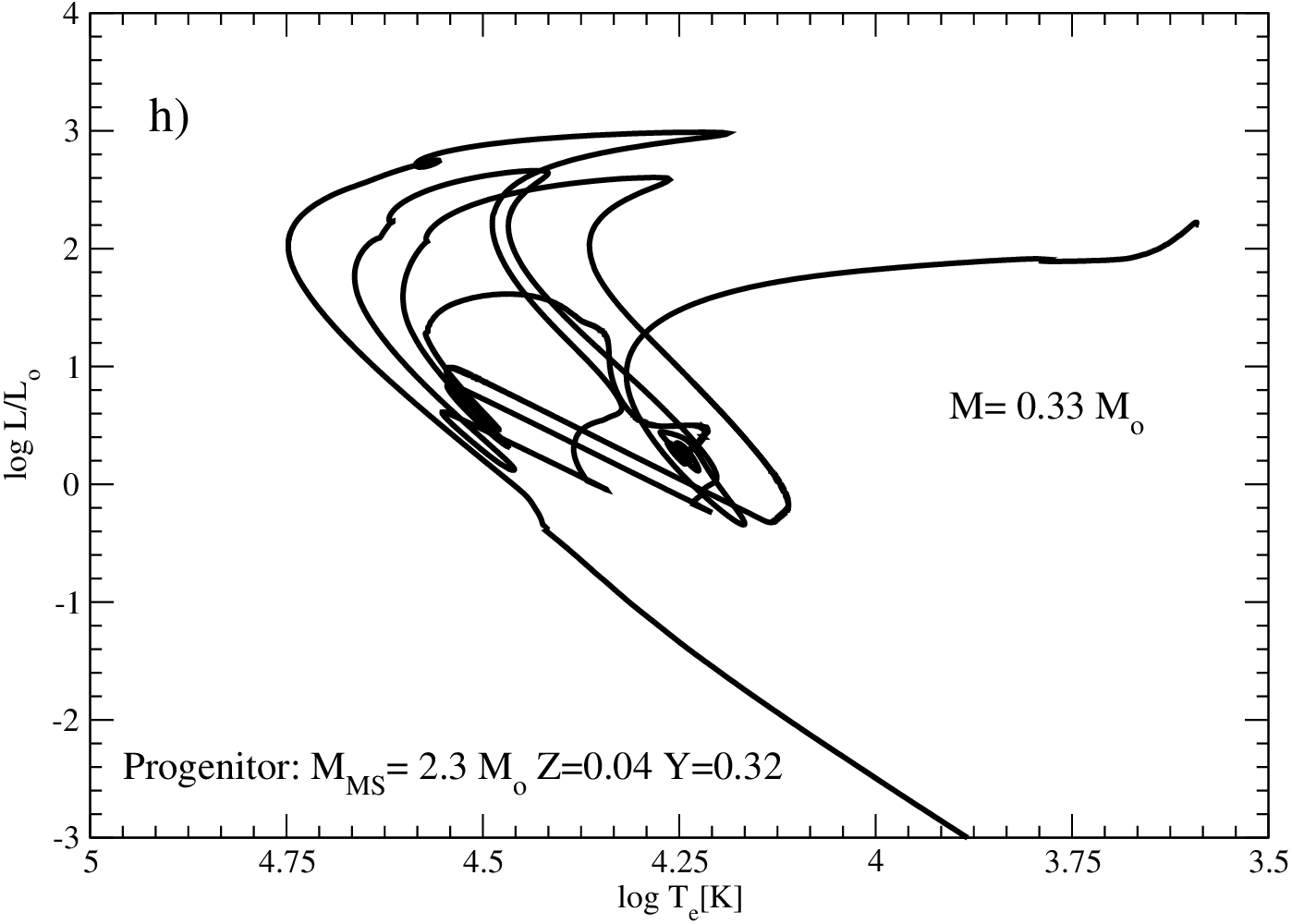}
     \end{minipage}
     \caption{Evolutionary tracks after the rapid mass loss episode occurred during the RGB 
     of the models described in the text. 
     The labelled masses are final massess or, more specfically, the residual 
     mass after the mass loss episode.
     In all the cases here illustrated, the progenitor is a star with M= 2.3 M$_{\odot}$ and Z=0.04.
     All the evolutionary models, 
     but the M=0.33 M$_\odot$ (panel $g$), produce a C-O WD at the end of the evolution.}
     \label{tracce}%
     \end{figure*}
In order to check such a working hypothesis and provide models of C-O WDs
 with mass lower than 0.5 M$_{\odot}$, we computed the evolution at constant
 mass of a star with M=2.3 M$_{\odot}$ until the red giant phase.
 Then at about $logL/L_{\odot}$=1.34, when M$_{H}$=0.257
 M$_{\odot}$, we switched on a mass loss regime at constant rate, 
 namely: 10$^{-7}$ M$_{\odot}$ yr$^{-1}$. Then,
we stopped this mass loss regime once required stellar masses was obtained and we 
followed the next evolution until the final cooling phase. This numerical recipe 
allowed us to compute a series of starting models of stars that underwent an
 intense mass loss episode on a time scale much shorter than the evolutionary 
time scale, as it happens during the Roche-lobe overflow in interacting binary stars or 
during the tidal stripping caused by close encounters between stars in 
crowded environment. The details of the rapid mass loss phase are not relevant 
for the following evolution. We have also checked that 
the overall evolution of the remnant do not sensibly depends on the luminosity at which 
the mass loss episode occurs, with the only exception of the smallest model, namely 
that of M= 0.33 M$_{\odot}$.

Figure \ref{tracce} shows the evolutionary tracks in the
HR-diagram for different values of the final mass, from 0.5 M$_{\odot}$ to
0.33 M$_{\odot}$, from the red giant to the final WD phase.
All the models showed in this figure, but that in
the bottom-left panel (i.e. panel $g$), succeed to ignite He and become
 WDs with a C-O core. The evolutionary tracks of the first three models,
 (those whose remnant mass is 0.5, 0.48 and 0.461 M$_{\odot}$) respectively, are quite similar. 
Once the mass loss has been switched off, the star continues to climb 
the red giant branch until the 3$\alpha$ reactions starts in the centre. The He
ignition occurs through a mild flash, which is much weaker 
than those found in lower mass stars (those developing high-degeneracy conditions),
 but stronger than that of a normal 2.3
M$_{\odot}$ star (no anomalous mass loss episodes). Note that the flash is stronger 
for the smallest remnant.  
After a few million years since the flash, a quiescent central He-burning
phase takes place in a convective core, while, as usual, the H-burning continues in a 
sorrounding shell. The star moves toward higher effective temperatures until 
the energy contribution of the 3$\alpha$ nuclear reaction to the total
 luminosity reached its relative maximum, then it turns-back to the red.
 The central helium exhaustion is followed by the onset of the He-burning in
 shell. At this point, the residual hydrogen-rich outer layer is thin, of the
 order of 0.06 and 0.03 M$_{\odot}$, for the model of 0.5 and 0.461
 M$_{\odot}$, respectively. The former model attempts to climb the asymptotic giant
branch until the thickness of the outer envelope is reduced down to approximately
0.01 M$_{\odot}$, when it starts to quickly move toward the blue. The latter
model, which starts the He-shell burning phase with a much thinner
hydrogen-rich envelope, leaves the red side of the HR diagram immediately after 
the end of the core He-burning. 
All the 3 models finally approach their WD tracks, where they cool down as a C-O WDs.      
  
From the analysis of these models emerges a general feature, which holds 
also for the models described in the following of the paper: the lower the mass of the remnant,
 the lower the mass fraction of the C-O core with respect to the total final mass.
  Consequently, the
 thickness of the helium-rich zone gets larger and larger as the mass of the
 WD decreases. 
The core of the 0.5 M$_{\odot}$ WD is about the 91\% of the total
mass\footnote{We define the C-O core as the region  where the helium mass fraction
is lower than than 0.5}, while those of the 0.48 and 0.461 M$_{\odot}$ are 
the 89\% and the 85\%, respectively.
%
    \begin{figure*}
     \vspace{0.7cm}
     \centering
     \begin{minipage}[c]{0.5\linewidth}
       \centering \includegraphics[width=0.9\linewidth]{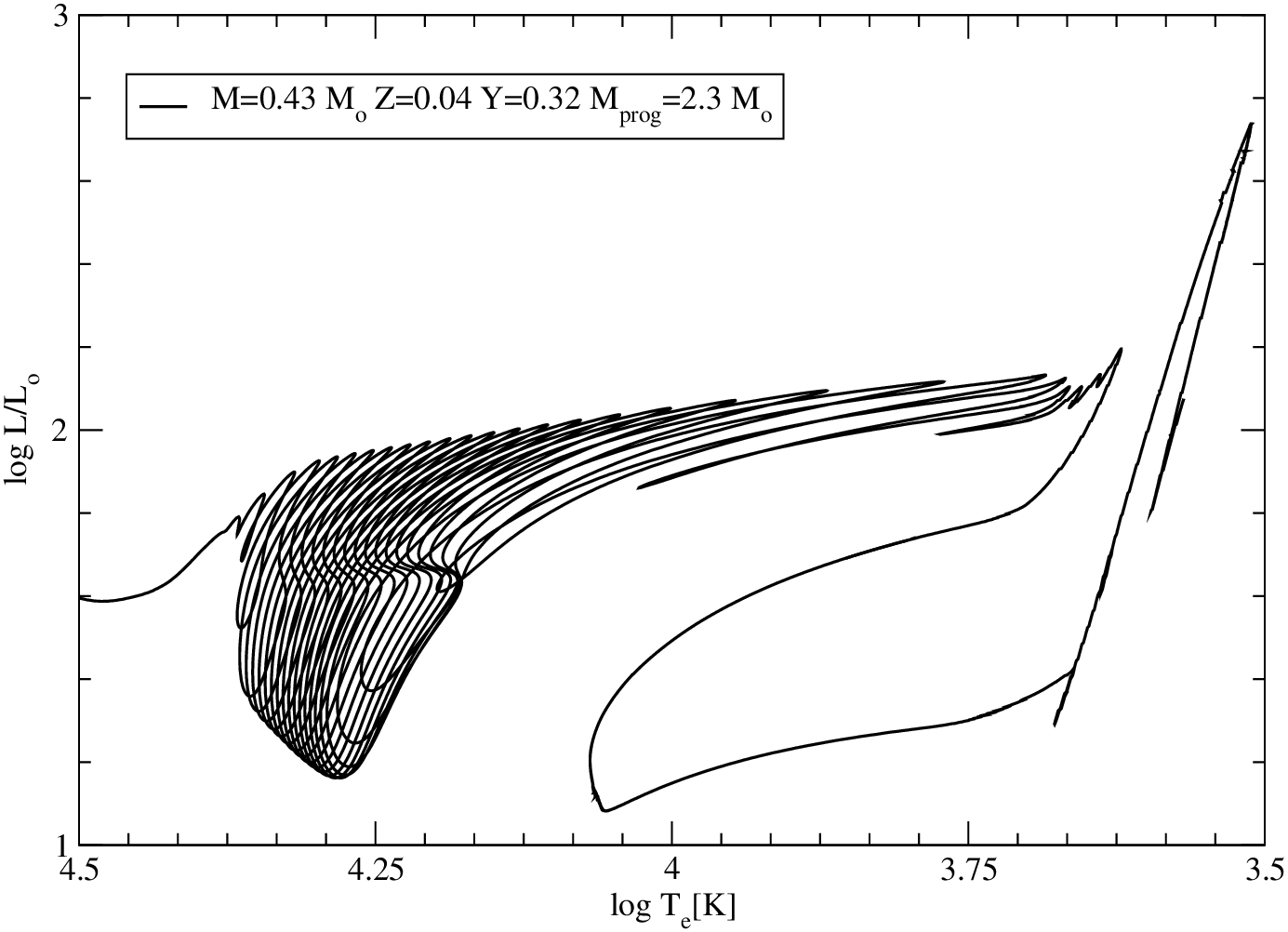}
     \end{minipage}%
     \begin{minipage}[c]{0.5\linewidth}
       \centering \includegraphics[width=0.9\linewidth]{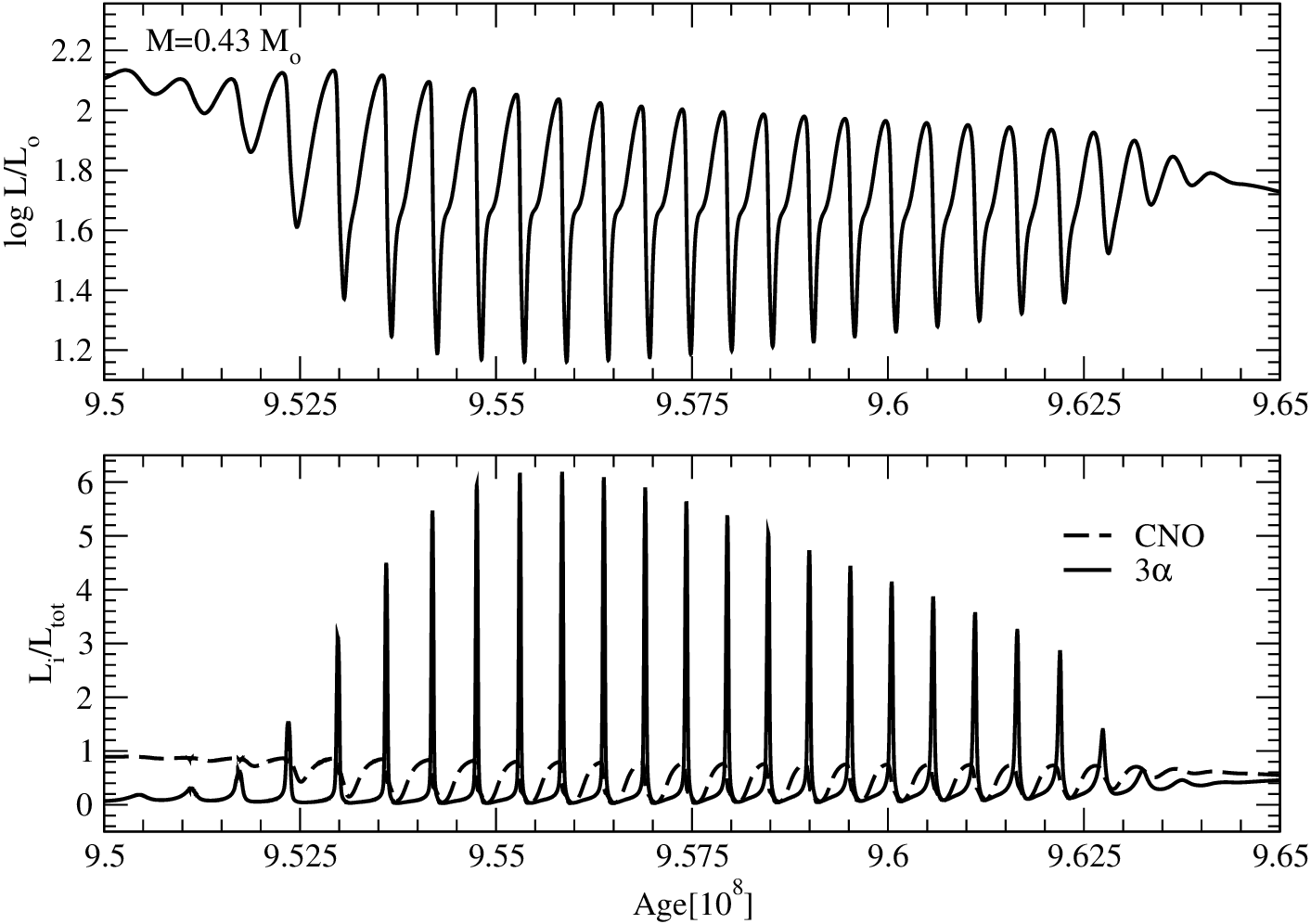}
     \end{minipage}
     \caption{Model of 0.43 M$_{\odot}$, whose porgenitor of M= 2.3 M$_{\odot}$
       underwent a strong mass loss episode during the red giant phase.
       Left panel: evolutionary track in the HR diagram from the RGB phase to
       the end of the He-thermal pulses. Upper-right panel: evolution  
       of the total luminosity during the thermal pulse phase. 
       Bottom-right panel: Contributions to the total luminosity of 
       the shell-H burning (dashed line) and the shell-He burning (solid line).}
     \label{0.43}%
     \end{figure*}
%

The evolution becomes much more complex and difficult to compute for lower
masses, as shown by the looping tracks of the models with remnant mass of
0.43, 0.4 and 0.38 M$_{\odot}$ plotted in figure \ref{tracce}
(panels $d$, $e$ and $f$). 
The evolution of the 0.43 M$_{\odot}$ star looks like those 
described above until the central-He exhaustion, when the model begins to move toward 
higher effective temperatures. At that point, after four pre-pulses 
marked by the small zig-zag in the HR diagram, the model      
 experiences a series of 20 He-shell thermal pulses: during each
flash the star describes a loop in the HR diagram. Such a series of 
thermally pulsing loops lasts about 10 Myr. 
Figure \ref{0.43} shows the evolutionary track of the 0.43 M$_{\odot}$ in the
HR diagram, zoomed around this phase, and the evolution as a
function of time of the total luminosity 
and the relative contributions of the shell-H burning and of the shell He-brning. 
After the last of these thermal
pulses, the model experiences a couple of post-pulses, again producing 
small zig-zags in the evolutionary track, followed by a quiescent phase during which the
shell He burning provides most of the energy,
after an initial short period during which the H shell was dominating.
In the meantime, the model moves toward higher effective tempearures
approaching the cooling track, where it experiences three strong
hydrogen-shell flashes. During these
episodes, the residual hydrogen-rich envelope is progressively eroded, until the 
star can cool down as a WD with a C-O core, whose mass is the 79\% of the total mass. 

The evolution of the model with 0.4 M$_{\odot}$ follows qualitatively the same
path of the 0.43 M$_{\odot}$ star, with the difference that, after the 
 thermally pulsing phase, with its characteristic loops in the HR diagram, it experiences a late
thermal pulse when the model is already approaching the cooling sequence, 
as shown in figure \ref{tracce} (panel $e$).
Finally, the star approaches again the cooling track and experiences
a few strong H-shell flashes.
The remnant is a WD with a C-O core whose mass is the 70\% of the total mass. 
%
    \begin{figure*}
     \vspace{0.7cm}
     \centering
     \begin{minipage}[c]{0.5\linewidth}
       \centering \includegraphics[width=0.9\linewidth]{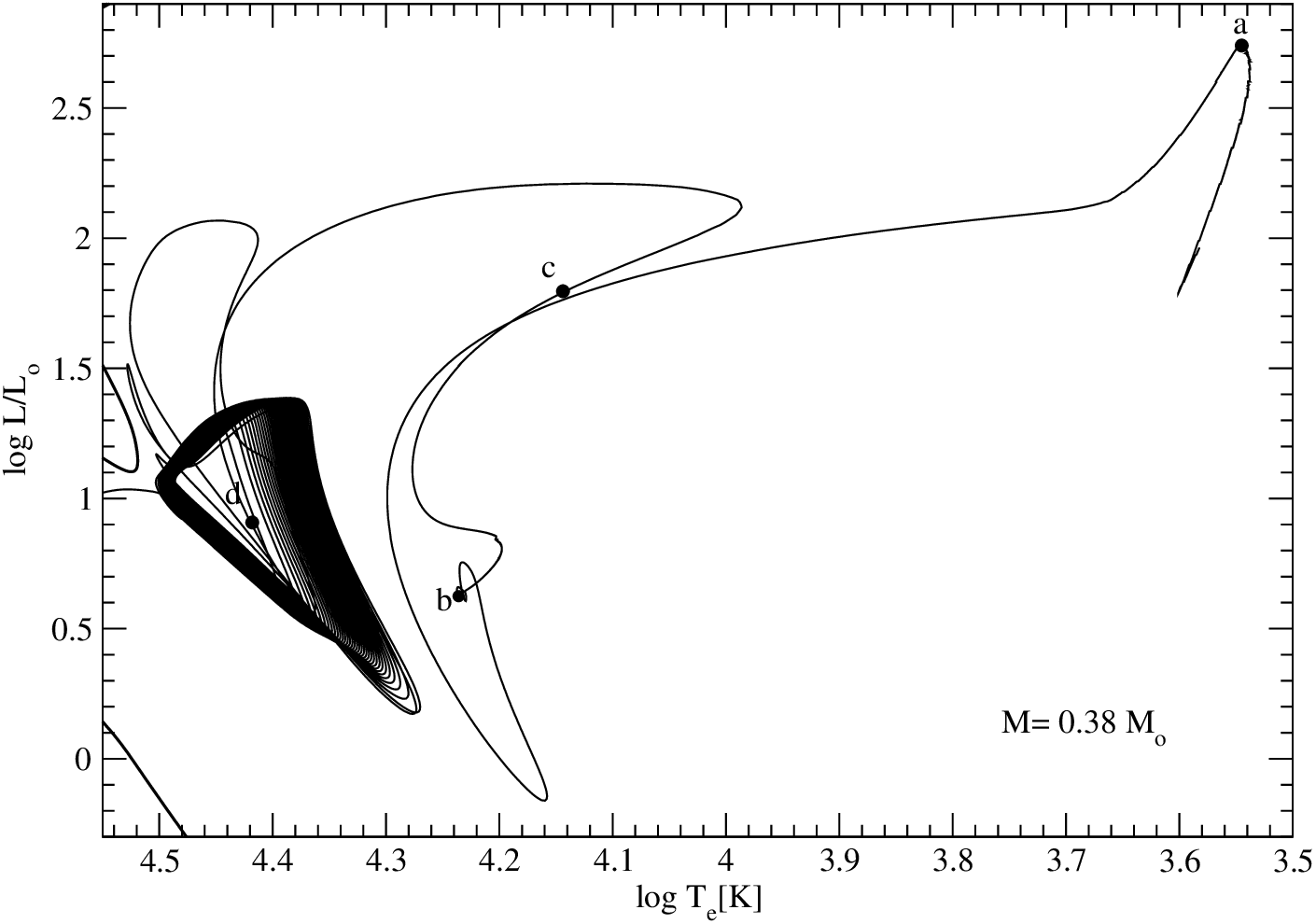}
     \end{minipage}%
     \begin{minipage}[c]{0.5\linewidth}
       \centering \includegraphics[width=0.9\linewidth]{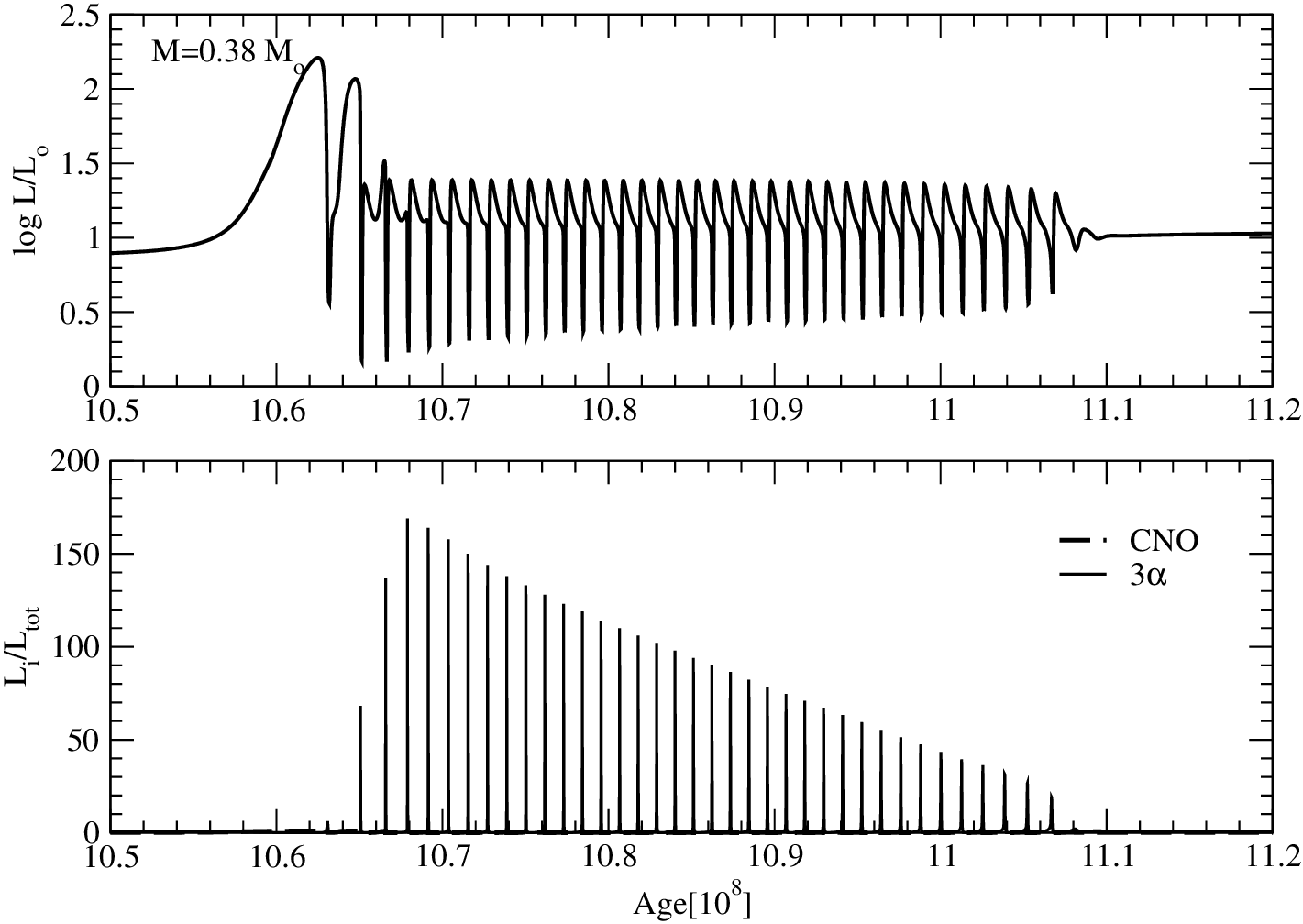}
     \end{minipage}
     \caption{Same as in figure \ref{0.43} but for the model of 0.38 M$_{\odot}$.
       The He-flash (point a), the beginning (point b) and the end (point c)
       of the quiescent He-burning in the convective core, the first of the two
       main He-thermal pulses (point d) are marked.}
     \label{0.38}%
     \end{figure*}
%

    \begin{figure}
     \vspace{0.7cm} 
     \centering
         \includegraphics[width=0.9\linewidth]{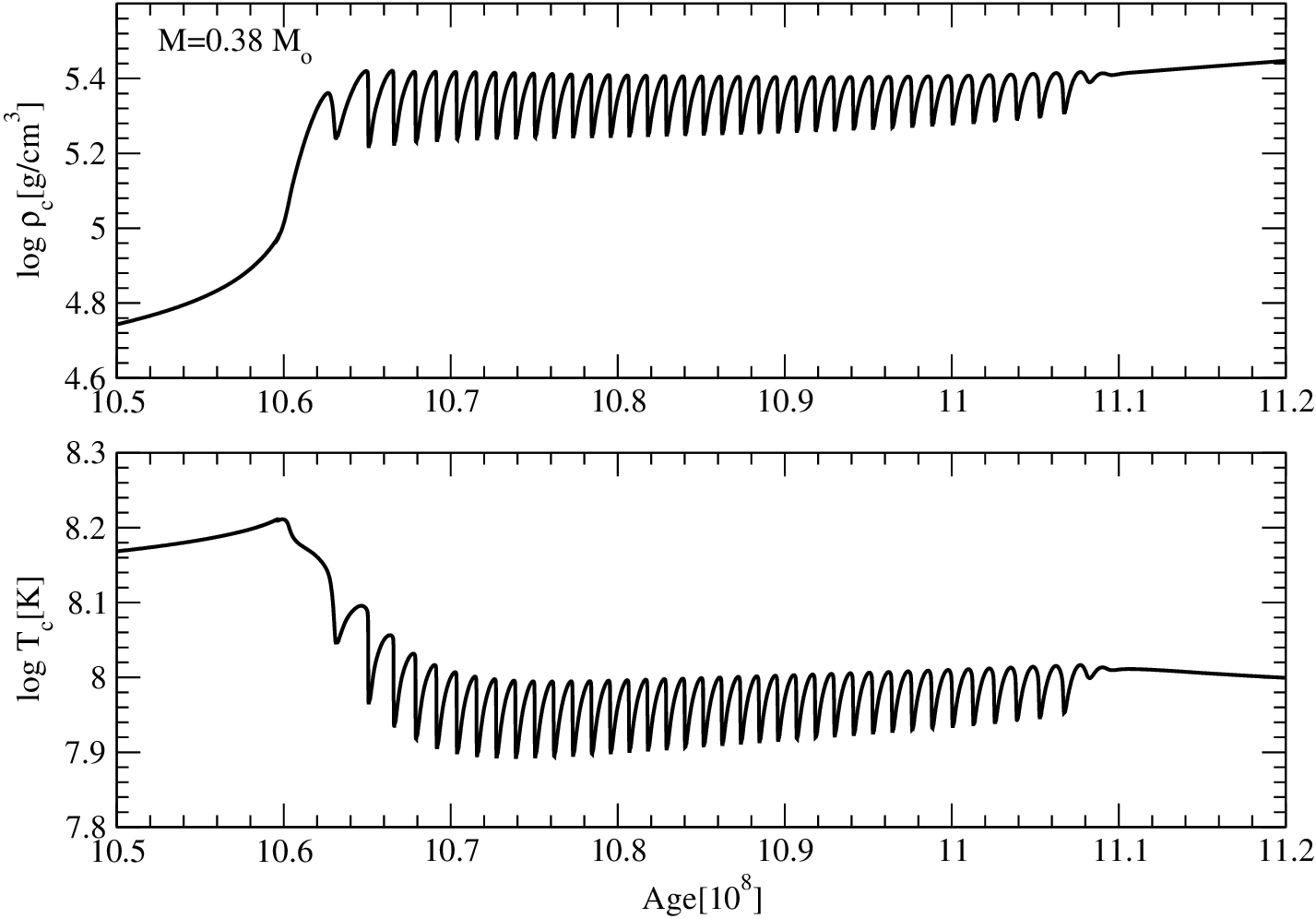}
     \caption{Model of 0.38 M$_{\odot}$. Evolution 
      of the central density (in g/cm$^3$, upper panel) and
       of the central temperature (in K, lower-panel) during the thermal
       pulse phase.  
       }
     \label{0.38-Tc}%
     \end{figure}

The evolution of the model with 0.38 M$_{\odot}$, shown in figure \ref{0.38},
 presents some peculiarities. 
The ignition of the 3$\alpha$ reaction occurs through a flash, when the star 
was already leaving the red giant branch (point $a$ in figure \ref{0.38}, right-panel). 
Such an He-flash, although substantially milder than that occurring in
low-mass stars, is stronger than those found for the models so far described. The
peak of the 3$\alpha$ luminosity is about 8.75 $\cdot 10^5$ L$_{\odot}$, to be compared to
that found in the normal 2.3 M$_{\odot}$ evolution (3.46 $\cdot 10^5$
L$_{\odot}$) or in the case of the model producing a remnant mass of 0.48 M$_{\odot}$ 
(5.24 $\cdot 10^5$ L$_{\odot}$).
The quiescent central He-burning evoltion began after about 5 Myr
 and proceeds in a convective core for 380 Myr (points $b$ and $c$ in figure
 \ref{0.38}). We are witnessing here an important feature of these peculiar
 objects, which we will discuss later in more detail: the quiescent central
He-burning lasts significantly longer than all other models.
Point $d$ in the figure 
marks the onset of the first of two main He-thermal pulses, followed by a series
of 38 thermal pulses lasting 46 Myr. Once again, during each of these pulses, the 
model describes a loop in the HR diagram.  
 Notice that the He-thermal pulses we 
showed above are not the same as that occur in standard AGB stars. 
In fact, in that case the thermal instability is the consequence
of the accumulation of a critical mass of helium accreted by the
 quiescent hydrogen-burning shell. 
While in this case the compression is due to the contraction of the core 
which follows the exhaustion of central helium. 
 To the best of our knowledge, Iben et
al. \cite{iben86} were the first to show and describe these
oscillations following the evolution of a remnant of 0.378 M$_{\odot}$ 
with a progenitor of 3 M$_{\odot}$. Subsequently, Bono et
al. \cite{bono97a, bono97b}, in a couple of papers devoted to the computation of evolutionary and
pulsational models of metal-rich stars, showed that the evolution after the
central helium exhaustion of models with reduced envelope is characterized by 
loops in the HR diagram, that they called gravonuclear loops. Since 
the quoted authors already fully and clearly described these gravonuclear
 instabilities, we will focus only on the their main features. 
As early understood in the pioneering study by Schwarzschild \& Harm
\cite{sh65}, the ignition of a nuclear reaction (i.e. the 3$\alpha$), 
whose efficiency sensitively depends on temperature, in a shell characterized
by a steep temperature profile leads to a thermal instability, because the
rate at which nuclear energy is released is larger than that at which is
dissipated by radiative transfer or converted in work to expand the outer
layers. As a consequence the temperature in the burning shell increases, hence
 the rate of nuclear energy release is further enhanced and a runaway onsets. 
This is the physical reason at the base of both the standard, and well-known,
 thermal pulses in AGB stars, and the peculiar, less-known, gravonuclear instabilities. 
On the other hand, what triggers the instability, i.e. the cause of the
compression of the He-shell and the consequent exceeding of the treshold
temperature for the nuclear reaction ignition, is different: the accumulation
of fresh helium produced by the H-burning shell in normal thermally pulsing
AGB stars and the contraction of the core after the central helium exhaustion
 in the present gravonuclear oscillations. 
Moreover, at variance with standard thermal pulses, during the loop
evolution the H- and He-shell never get near each other and, eventually, the He-burning shell
is quenched by the expansion of the envelope.
Figure \ref{0.38-Tc} shows the time evolution of the central density (upper
panel) and the central temperature (bottom panel) of the 0.38 M$_{\odot}$
model during the gravonuclear oscillations. As early shown by Iben et
al. \cite{iben86} and  Bono et al. \cite{bono97a, bono97b}, at variance with standard thermally pulsing phase
 in normal AGB stars, these pulses affect the entire star, with large
 oscillations in the central quantities, of the order of 40\% in density and
 20\% in temperature.
After the last of these thermal pulses the model gets near to 
the WD sequence, where it experiences four strong H-flashes and, 
finally, cools down as WD whose C-O core is only the 50\% of the total mass.

Figure \ref{tracce} (panels $g$ and $h$) shows two different evolutionary tracks, corresponding to the same final mass
 (M= 0.33 M$_{\odot}$), but leading to the formation of WDs with different composition.
In the former case (panel $g$ of figure \ref{tracce}), the rapid mass loss
episode starts at the same point of the RGB evolution as in all previous cases.
This model fails to ignite He in the core and becomes a He WD. 
In the second model, the rapid mass loss episode on the RGB has been turned on later, 
when the mass of the H-exhausted core is $\approx$ 1\% more massive than in all previous cases.
In such a case, the star enters the core-He burning phase 
(panel $h$ of figure \ref{tracce}).
When the star is already leaving the RGB, the He ignition occurs
through a mild flash. 
 During the flash, an extended convective core (about 80\% of the total mass) suddenly appears. Notice the
 quick decrease of the surface luminosity, due to the temporary stop of the H-burning shell caused by
 the expansion and the consequent cooling powered by the energetic outcome of the He burning.
Then, for a certain time, an oscillation 
of the He-burning luminosity takes place, correlated to an oscillation of the extension of the convective 
core and anti-correlated to an oscillation of the gravitational energy release.   
Meanwhile, the evolutionary track describes small loops in the HR diagram. 
Later on, a long-lasting quiescent He-burning sets in a convective core, whose maximum mass is about 
0.1 M$_{\odot}$. Once the central He is nearly ehausted, the H-burning shell resumes and becomes
the dominant energy source.  
Approaching the WD sequence, the model
experiences three strong H flashes, corresponding to the large loops 
in the HR diagram. Finally, the star cool down as a WD
 with a C-O core of about the 53\% of the total mass. 
This is the lowest C-O WD we managed to produce for this chemical composition,
since models with mass smaller than 0.33 M$_{\odot}$ do not ignite 
the 3$\alpha$ reactions and terminate their evolution as He-core WDs.
%
    \begin{figure}
     \vspace{0.7cm}
     \centering
        \includegraphics[width=0.85\linewidth]{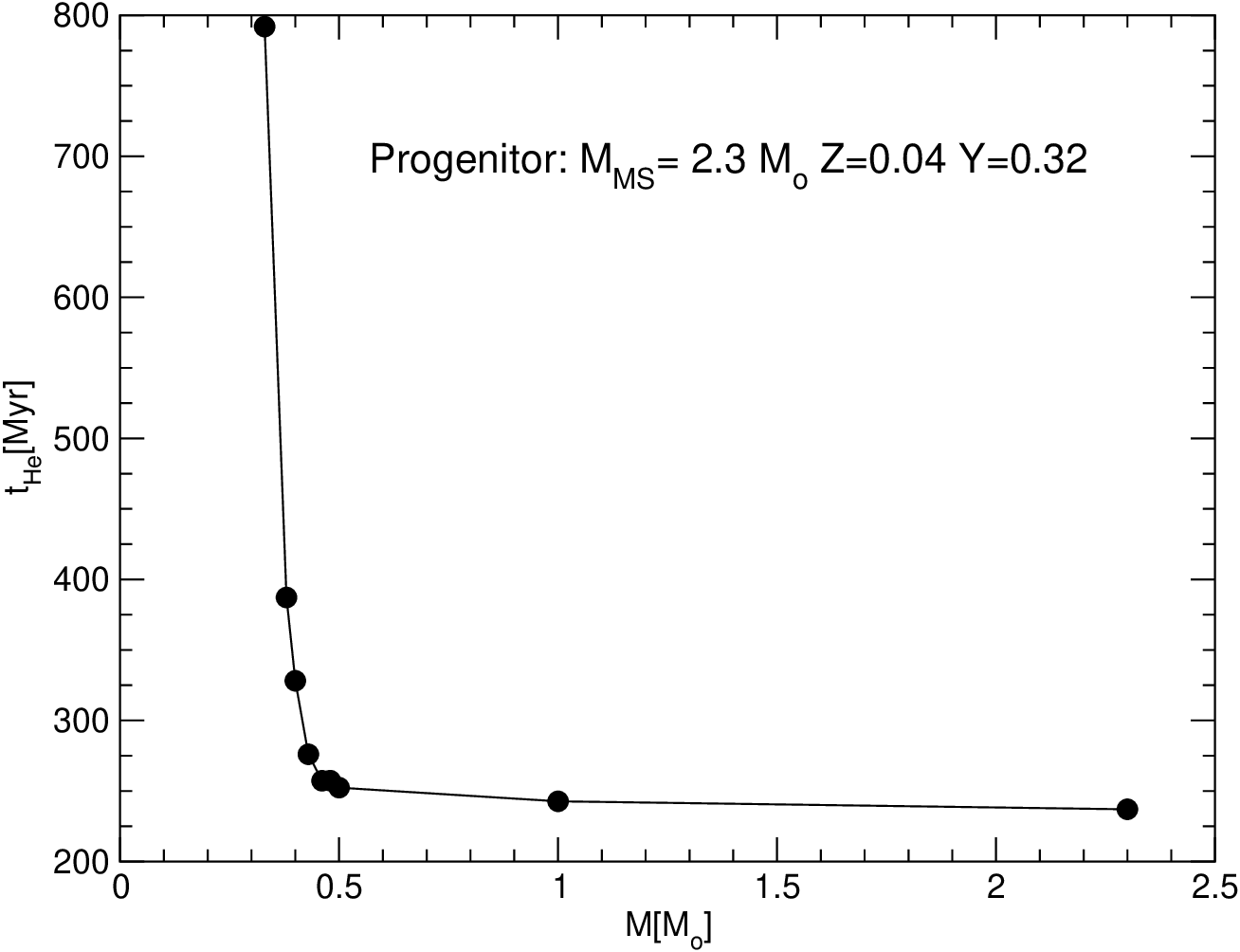}
     \caption{Helium burning lifetime (in Myr) as a function of the total mass
       (in  M$_{\odot}$).
       }
     \label{The_M}%
     \end{figure}

Let us stress two peculiar properties of the C-O VLMWDs.
Figure \ref{The_M} shows the duration of the central He-burning $t_{He}$ 
as a function of the total mass of the remnant obtained by stripping part of the envelope
mass along the RGB of a progenitor star with initial mass 2.3 M$_{\odot}$, whose
evolution has been discussed above. 
As already noted, the smaller the remnant mass the larger the 
time spent by the models in the central He-burning
phase. Such an increase is quite smooth and shallow for remnant mass larger
than 0.5 M$_{\odot}$, but becomes
very steep for smaller masses. The smallest models which succeeds to 
ignite He-burning, that is the M= 0.33 M$_{\odot}$, takes about 800 Myr to 
exhausts its central helium, which is more than three time longer  
than the value of the standard 2.3 M$_{\odot}$ star. Note that for this peculiar 
star, the core He-burning lasts more than the central-H burning of the 
progenitor ($\tau_H \approx$ 613 Myr). It is, by far, the longest core-He burning lifetime.
As previously recalled, the He-burning lifetime depends on the 
stellar luminosity, which is smaller for star with smaller H-exhausted core at the He ignition.
Thus, looking at the core masses plotted in figure 1, it results that
the longest He-burning lifetime for standard models is attained by 
the star with about 2.3 M$_\odot$, the one corresponding to the minimum 
of the RGB fase transition (Castellani, Chieffi, Straniero
\cite{ccs92}; Castellani et al. \cite{cast2000}; Dominguez et
al. \cite{dominguez}; Girardi \cite{girardi}).
Table \ref{tabMhe} lists the mass of the H-exhausted core M$_{H}$, taken  
at the time when the central He abundance decreased of 0.001 from the
initial value, for models of different total mass M. As one can see, 
  at the He ignition, all the models here presented have, more or less,
 the same core mass of the standard 2.3 M$_\odot$ model. 
%
\begin{table}
\begin{minipage}[t]{\columnwidth}
\caption{He-burning ignition.}             
\label{tabMhe}     
\centering                 
\begin{tabular}{c c}       
\hline\hline                
M[M$_{\odot}$]\footnote{stellar mass} & M$_{H}$[M$_{\odot}$]\footnote{mass of
  the H-exhausted core.
  As a reference evolutionary point we chose the model in which the
  central He abundance decreased of 0.001 from the initial value.} \\   
\hline                       
0.330   &     0.323  \\   
0.380   &     0.327  \\    
0.400   &     0.325  \\   
0.430   &     0.326  \\  
0.461   &     0.326  \\   
0.480   &     0.326  \\  
0.500   &     0.325  \\     
2.300   &     0.323  \\   
\hline                                 
\end{tabular}
\end{minipage}
\end{table}
%
 Nonetheless, the substantial erosion of the H-rich envelope causes a 
depression of the shell-H burning that is more evident for models producing smaller remnants. 
 As a consequence, the rate of growth of the M$_H$ core during the central
  He-burning, and thus the luminosity evolution and the length of this phase,
 is quite different in remnants with different envelope
  thicknesses. The lower the mass of the remnant, the less efficient the H-burning
  shell and the slower the increase of the M$_H$ core. In the extreme case of the 0.33
M$_{\odot}$ model, the mass of the H-exhausted core remains almost constant
during the entire central He-burning, as the H-shell is pratically turned
off. This explains the steep growth of the central He-burning lifetime $t_{He}$ as
the mass of the remnants decreases below 0.5  M$_\odot$.
Note that, due to the very long He-burning lifetime, the age of the 0.33
M$_{\odot}$ model at the beginning of the cooling sequence is 1.5 Gyr,
to be compared with the 0.9 Gyr of the standar 2.3 M$_{\odot}$ model, 
which is expected to produce a C-O WD of about 0.6 M$_\odot$.

\begin{figure}
     \vspace{0.7cm}
     \centering
          \includegraphics[width=0.9\linewidth]{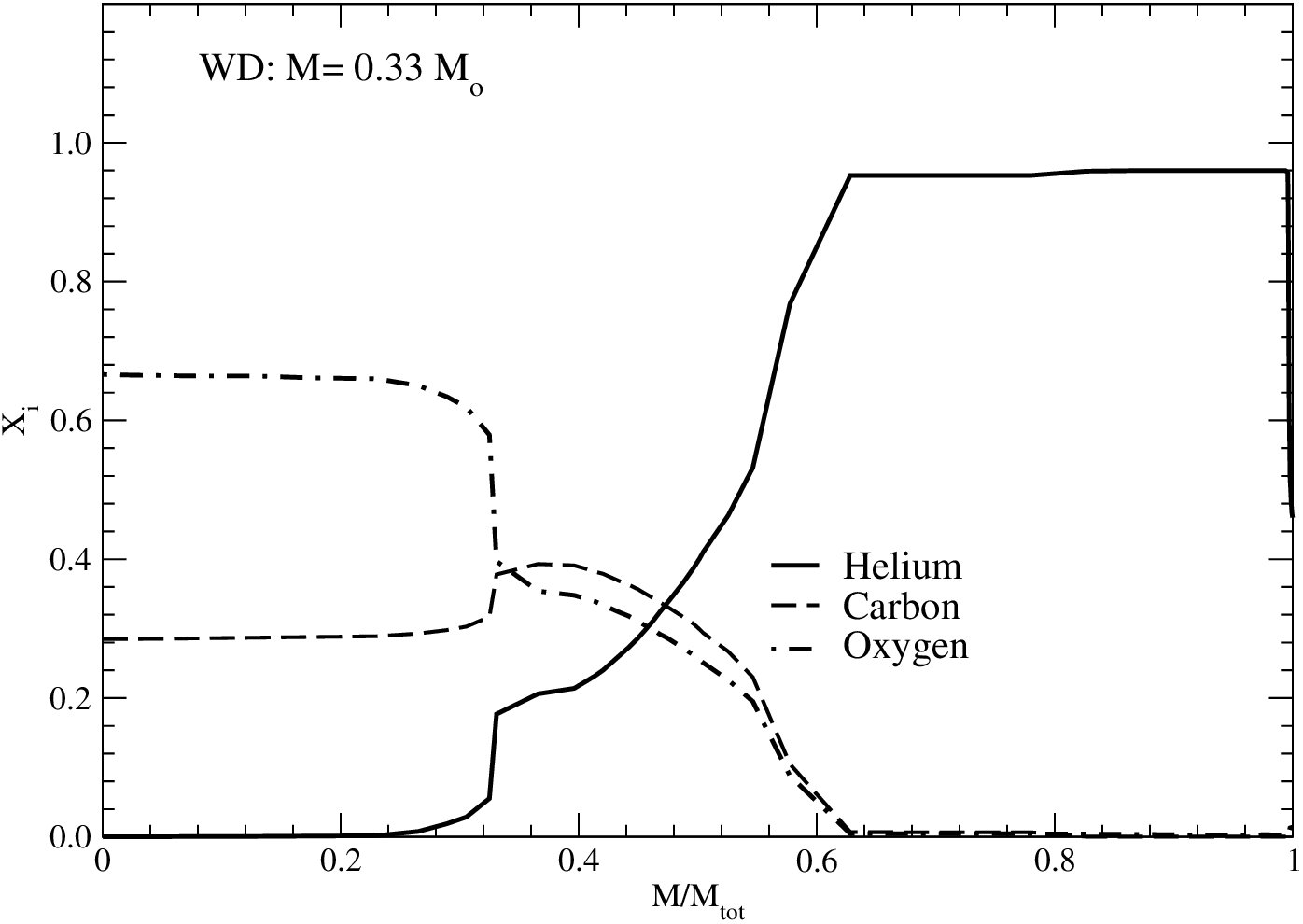}
     \caption{Internal mass fraction of helium (solid line), carbon (dashed line) and oxygen 
     (dot-dashed line) as a function of the mass coordinate for the C-O WD with M= 0.33 M$_{\odot}$.
       }
     \label{profili}%
     \end{figure}

Another interesting feature of the C-O VLMWDs concerns their internal composition.
Figure \ref{profili}
shows the chemical abundance profiles of $^4$He, $^{12}$C and $^{16}$O of the 
C-O WD of 0.33 M$_{\odot}$.
As previously stated, the lower the mass of the remnant, the lower the fraction of the total mass 
confined in to the C-O core or, equivalently, the larger the fraction of mass confined in to 
the helium-rich external layer. 
In the case of the  0.33 M$_{\odot}$ model, the He-rich layer 
is about the 50\% of the total mass, that is very different from the value 
found for the more massive (normal) C-O WDs, namely 1-2 \%.

\section{Cooling evolution}
%

%
    \begin{figure}
        \vspace{0.7cm} 
     \centering
         \includegraphics[width=0.9\linewidth]{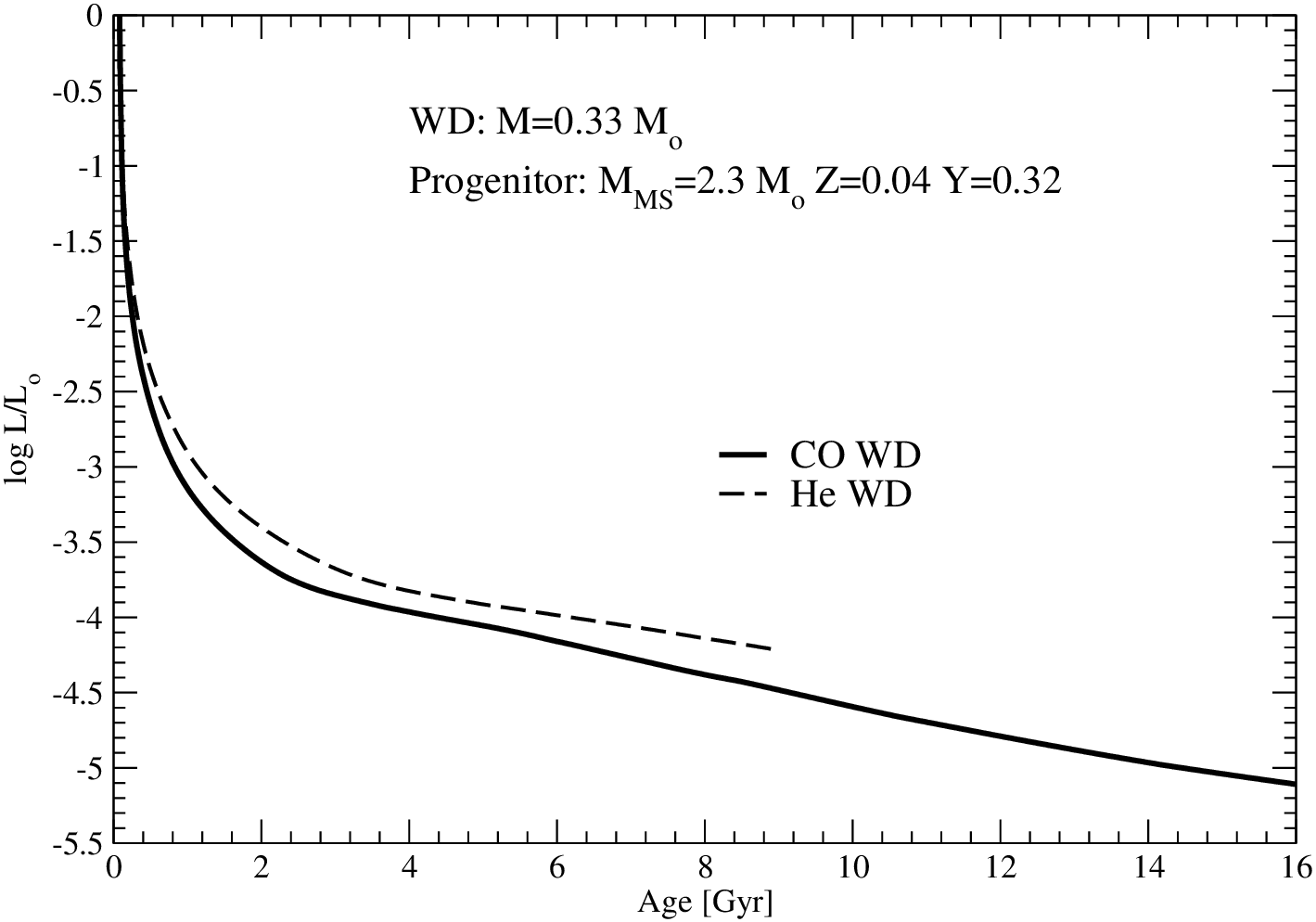}
     \caption{Comparison between the cooling curves,
      logL/L$_{\odot}$ vs. time, of the two WDs of 0.33 M$_{\odot}$,
      that with the He-core (dashed line) and with C-O core (solid line).}
     \label{cooling}%
     \end{figure}
    \begin{figure}
    \vspace{0.7cm}
    \centering 
        \includegraphics[width=0.9\linewidth]{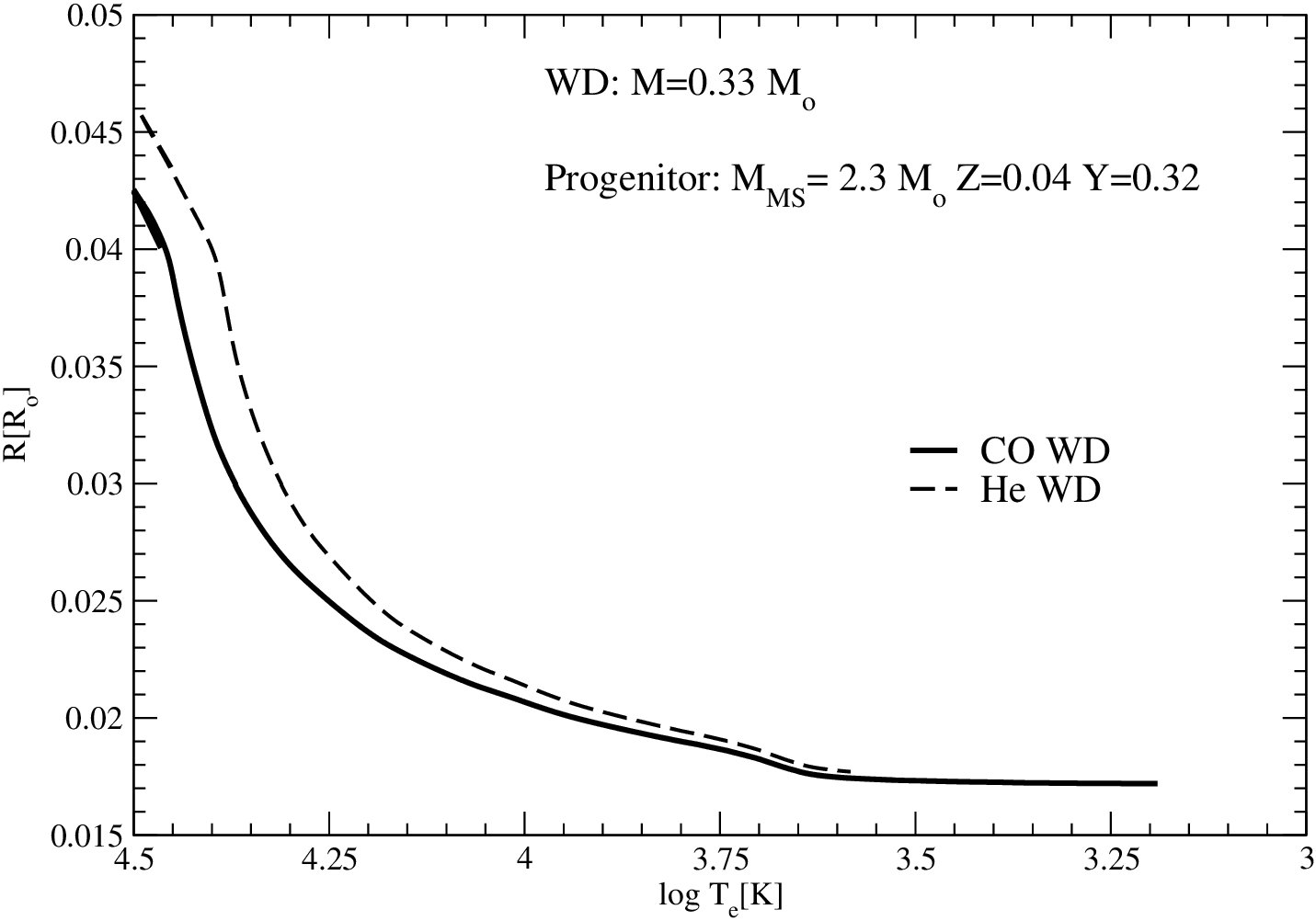}
    \caption{Radius vs. effective temperature of the two WDs of
             0.33 M$_{\odot}$, that with the He-core (dashed line)
             and with C-O core (solid line).}
     \label{coolingRTe}%
     \end{figure}
The present evolutionary computations confirm that it is possible 
to have C-O WDs with mass as low as 0.33 M$_{\odot}$, significantly lower
than 0.5 M$_{\odot}$, the classical and commonly accepted lower limit. 
This means that in the mass range 0.33 - 0.5 M$_{\odot}$ both He and C-O core 
WDs can exist.
We will focus on the comparison between the structure and evolution of the two 
remnants of 0.33 M$_{\odot}$ with different core compositions. 

The thickness of the H-rich outermost layer of the two models of 0.33 M$_\odot$ is 
practically the same, namely M= 0.0014 M$_{WD}$ for
 the He WD and M= 0.0015 M$_{WD}$ for the  C-O WD.
Figure \ref{cooling} shows the comparison between the cooling curves
of the He-core WD (dashed line) and the C-O core (solid line). 
As expected (see the detailed discussion by Panei et al. \cite{panei}),
 the remnant with the He-rich core cools slower
 than that with a C-O core, as due to the larger specific heat,
so that the thermal content of the He WD is larger than that of the C-O one with the same total mass.
Such a difference in the cooling times is partially counterbalanced 
at $logL/L_{\odot} \approx$ -4, when the C-O core begins to crystallize.

Figure \ref{coolingRTe} shows the comparison between the radii as a function
 of the effective temperature for the two WDs of 0.33 M$_{\odot}$. 
For a given effective temperature, the He-core WD is more expanded
 than the other one, the difference beeing not negligible at the beginning of
 the cooling ($\sim$ 8.5\%). In principle, this difference of the radii 
at a given effective temperature 
offers the possibility to distinguish the core composition of WDs in the 
mass range 0.33 - 0.5 M$_{\odot}$. 
%
    \begin{figure}
    \vspace{0.7cm} 
     \centering
          \includegraphics[width=0.9\linewidth]{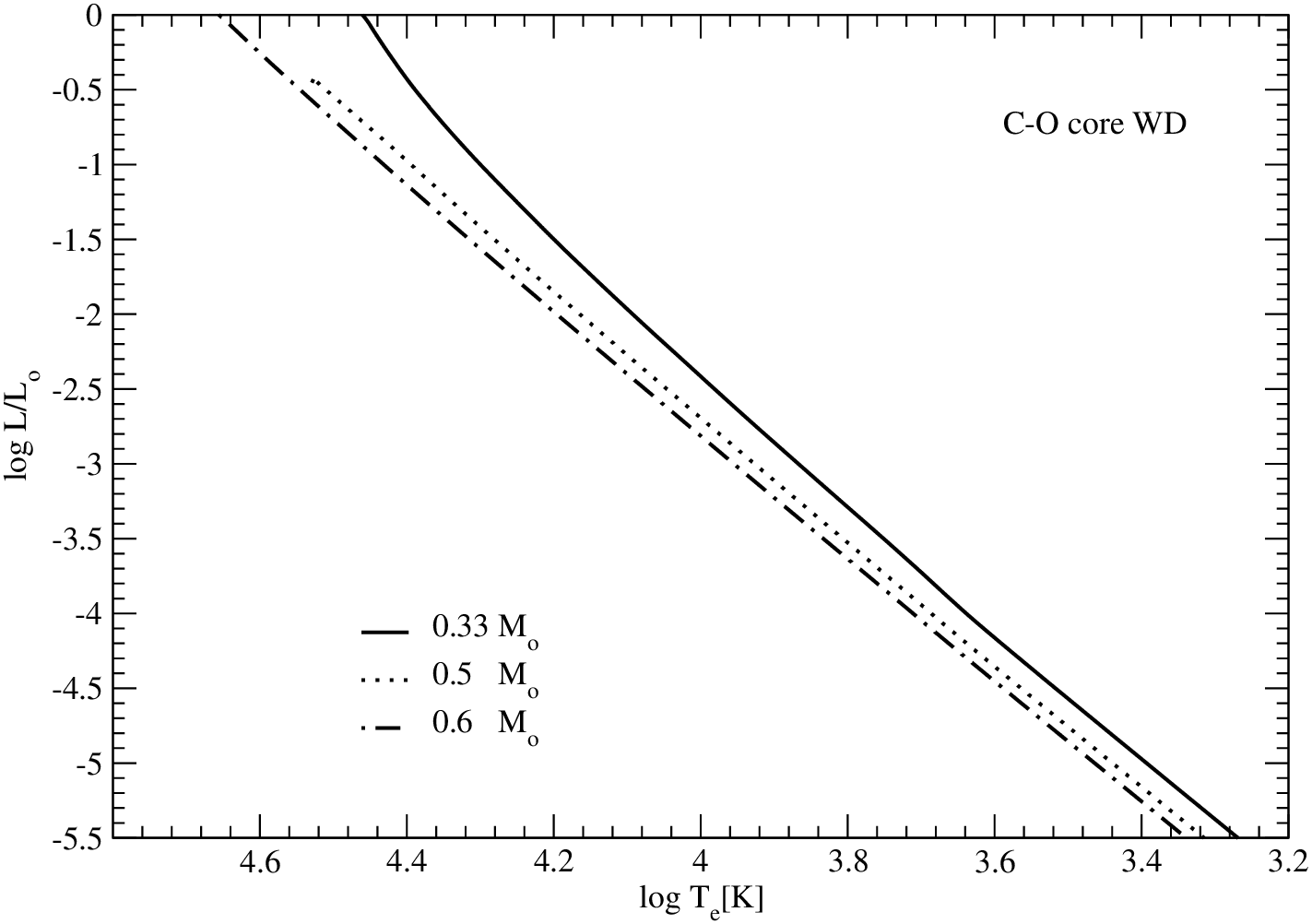}
     \caption{Comparison between the cooling tracks of C-O core WDs of masses:
       0.33 M$_{\odot}$ (solid line), 0.5 M$_{\odot}$ (dotted line), 0.6 M$_{\odot}$ (dot-dashed line).
       }
     \label{trackWD}%
     \end{figure}
    \begin{figure}
     \vspace{0.7cm}
     \centering
          \includegraphics[width=0.9\linewidth]{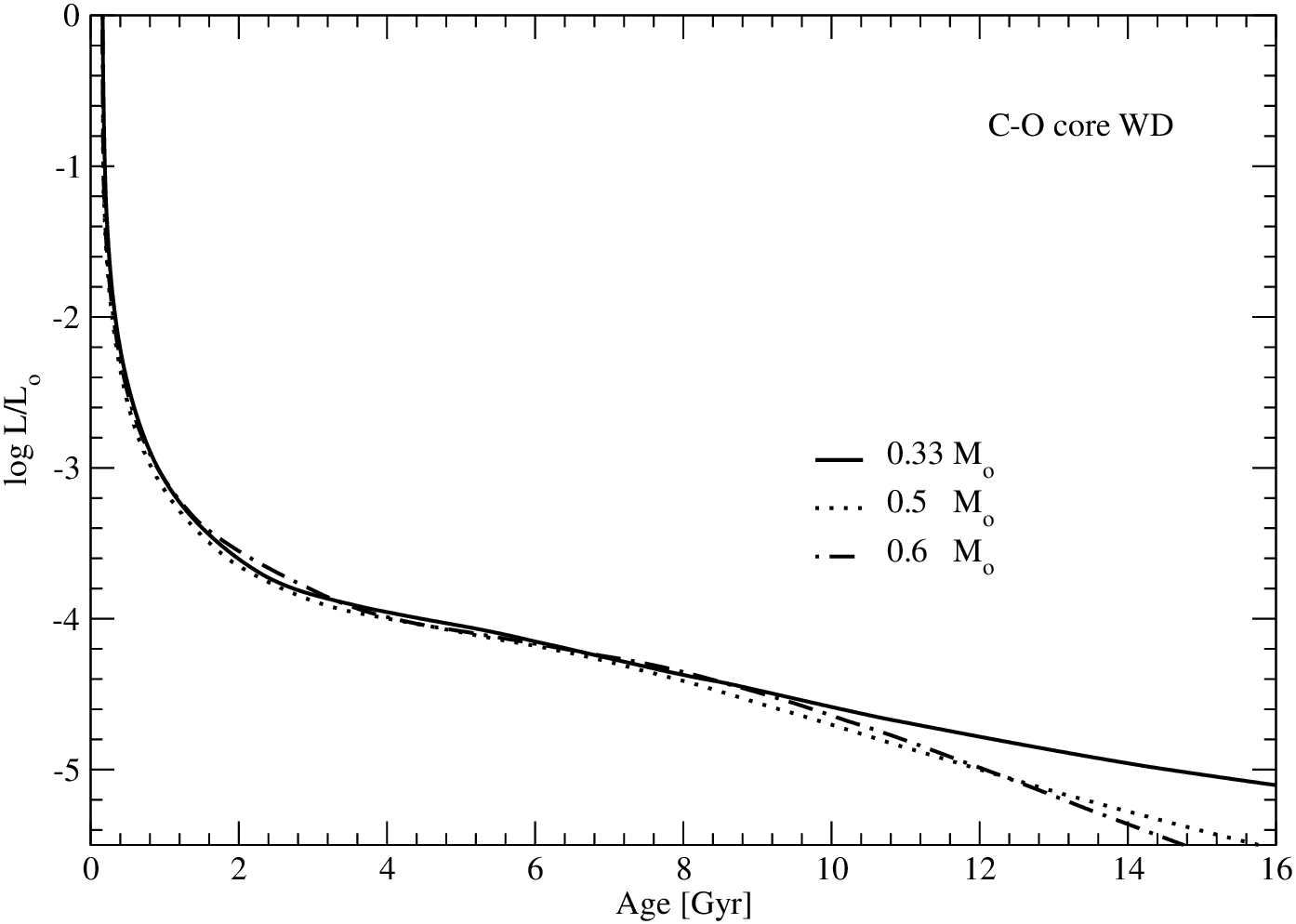}
     \caption{Comparison between the cooling curves, $logL/L_{\odot}$
       vs. time, of C-O core WDs of masses: 0.33 M$_{\odot}$ (solid line), 0.5
       M$_{\odot}$ (dotted line), 0.6 M$_{\odot}$ (dot-dashed line).
       }
     \label{cooling06}%
     \end{figure}
Since the C-O VLMWDs are the result of non-standard evolutionary channels,
it might be of some interest to compare their main characteristics 
to those of the C-O WDs produced by the evolution of non-interacting-single stars 
(i.e. WD with mass M$\ge$ 0.5 M$_{\odot}$). Figures 
\ref{trackWD} and \ref{cooling06} show the comparison between the tracks in the HR
diagram and the cooling curves, respectively, of the lowest C-O
core WD we managed to build, namely that of 0.33 M$_{\odot}$, with the more standard 0.5
and 0.6 M$_{\odot}$ remnants. As it is expected, the evolutionary tracks of the low mass C-O WD is
redder than those of the more massive objects and marks the reddest frontier of the
C-O WD loci. The tracks of these peculiar objects overlap the region of the HR diagram 
occupied by the He WDs.
Concerning the comparison among the cooling curves, as shown in figure
\ref{cooling06}, the luminosity evolutions look like quite similar, although not
identical, down to $logL/L_{\odot} \approx$ -4.4. Then, at fainter magnitudes or for cooling ages
larger than $\approx$ 9 Gyr,
the evolution of the standard WDs get faster and faster and their luminosity
drop more quickly than that of the 0.33 M$_{\odot}$ model. When the cooling age is 
14 Gyr, the 0.6 M$_{\odot}$ WD reaches
$logL/L_{\odot} \approx$ -5.4, while the 0.33 M$_{\odot}$ one is still at -4.9. The 
reason for such a behavior is that the solid core of the VLMWD is less dense than those of the other two objects
\footnote{the central density of the 0.6 WD is $\approx 5$ times larger than that of the 0.33 WD},
an occurrence causing a delay of the onset of the Debye-cooling regime.
This difference in the luminosity becomes even more large when the longer lifetime 
of the progenitors of the VLMWDs is taken in to account. 

\section{Summary and conclusions}
By means of fully consistent evolutionary computations, we proved 
 that the minimum mass for a C-O WD is much lower than the commonly agreed 0.5
 M$_{\odot}$. In fact, we described the evolutionary paths leading to the production of C-O WDs
 with mass as low as 
0.33 M$_{\odot}$, nearly the value of the minimum  
 M$_H$ required for the He ignition in non-degenerate conditions, which
     depends only slightly on the chemical composition. 
 These VLMWDs with a C-O core might result from the evolution of a star with initial mass
around 2.3 M$_{\odot}$, which undergoes a considerable mass loss along the red
giant phase.

As a consequence, in the mass range 0.33-0.5 M$_{\odot}$ both He and C-O core
 WDs can exist. 
As expected and already shown by the detailed analysis of Panei et al. \cite{panei},
 the cooling times of these two classes of WDs are quite different, being the
 He-core remnants significantly slower than the C-O ones. Furthermore, we
 showed also that the He WDs are more expanded that their C-O counterparts at
 a given effective temperature. In principle, such an occurrence would allow to discriminate 
between WDs with an He-core and those with a C-O-core in the mass range where both 
remnants exist, namely 0.33-0.5 M$_{\odot}$. 

Large samples of observed VLMWDs (M$<$0.5 M$_{\odot}$), belonging to the field
 (Liebert et al. \cite{liebert}, Maxted et al. \cite{maxted},
 Eisenstein et al. \cite{eisenstein}) or to stellar clusters 
(Bedin et al. \cite{bedin05}, Monelli et al. \cite{monelli}, Kalirai et al. \cite{kalirai}, 
Bedin et al. \cite{bedin08}, Calamida et al. \cite{bedin08}) are now available.
In the next future, basing on the models here discussed, it would be  possible to identify 
the observational counterpart of very low mass remnants with a C-O core among those
 commonly ascribed to the He-core WD population.  
 
The computation of the evolution of the stripped progeny of the 2.3
M$_{\odot}$ provided also the evidence for the longest lasting central
He-burning stellar objects. In fact, we showed that the smaller the remnant
 mass, the longer the central He-burning phase. Such an increase becomes
very steep for masses lower than 0.5 M$_{\odot}$ and reaches the maximum for the M= 0.33
M$_{\odot}$ model, which takes about 800 Myr to 
exhausts its central helium, which is more than three time longer  
than the time required by the longest lasting He-burning standard star,
i.e. the 2.3 M$_{\odot}$. Thus, the M= 0.33 M$_{\odot}$ model attains the 
longest core He-burning lifetime. 

\begin{acknowledgements}
It's a pleasure to thank Giuseppe Bono and Scilla Degl'Innocenti, who kindly read the paper, for
the many useful and pleasant discussions and the referee (Han Zhanwen) for the
positive and useful report. PGPM has been supported by PRIN-MIUR
2007 ({\em Multiple stellar populations in globular clusters: census,
  characterizations and origin}, PI G. Piotto), and OS by the PRIN-INAF program 2008.
\end{acknowledgements}

\end{document}